\documentclass[acmsmall,nonacm]{acmart}

\usepackage{makecell}
\usepackage{bm}
\usepackage{enumitem}
\setlist{nolistsep}
\usepackage{multirow}
\usepackage{float}
\usepackage{tabularx}
\usepackage{pifont}
\usepackage{booktabs}
\usepackage{colortbl}
\usepackage{mathtools}
\usepackage{caption}
\usepackage[subrefformat=parens,skip=0pt]{subcaption}

\usepackage[detect-all]{siunitx}
\usepackage{xspace}
\usepackage[bb=boondox,bbscaled=.95,cal=boondoxo]{mathalfa}
\usepackage[%
	sort&compress
]{cleveref}
\crefname{figure}{Fig.}{Figs.}
\Crefname{appendix}{Supplement}{Supplements}

\newcommand\ie{i.e.\xspace}
\newcommand\eg{e.g.\xspace}

\newcommand\US{U.S.\xspace}

\newcommand{\X}{{X (formerly Twitter)}\xspace}

\newcommand{\var}[1]{\mathit{#1}}

\DeclareUnicodeCharacter{2212}{-}

\AtBeginDocument{%
  }

\setcopyright{cc}
\copyrightyear{2026}
\acmYear{2026}
\acmDOI{XXXXXXX.XXXXXXX}
\acmConference[Conference acronym 'XX]{Make sure to enter the correct
  conference title from your rights confirmation email}{June 03--05,
  2018}{Woodstock, NY}
\acmISBN{978-1-4503-XXXX-X/2018/06}

\begin{document}

\title{Hiding Liker Identity Did Not Increase Engagement With Reputationally Risky Content on X (Formerly Twitter)}

\author{Yuwei Chuai}
\affiliation{%
  \institution{University of Luxembourg}
  \city{Luxembourg}
  \country{Luxembourg}
}
\email{yuwei.chuai@uni.lu}

\author{Manoel Horta Ribeiro}
\affiliation{%
  \institution{Princeton University}
  \city{Princeton}
  \state{New Jersey}
  \country{United States}
}
\email{manoel@cs.princeton.edu}

\author{Gabriele Lenzini}
\affiliation{
  \institution{University of Luxembourg}
  \city{Luxembourg}
  \country{Luxembourg}}
\email{gabriele.lenzini@uni.lu}

\author{Nicolas Pr{\"o}llochs}
\affiliation{
 \institution{JLU Giessen}
 \city{Giessen}
 \country{Germany}}
\email{nicolas.proellochs@wi.jlug.de}


\begin{abstract}
In June 2024, X (formerly Twitter) made likes from public to private, offering a rare, platform-level opportunity to study how the visibility of engagement signals affects users' behavior. Here, we investigate whether hiding liker identities increases the number of likes received by high-reputational-risk content, content for which public endorsement may carry high social or reputational costs due to its topic (e.g., politics) or the account context in which it appears (e.g., partisan accounts). To this end, we conduct two complementary studies: 1) a Difference-in-Differences analysis of \num{153704} posts that are created by \num{1045} accounts and have received over 324 million likes on X (formerly Twitter) before and after the policy change; 2) a within-subject survey experiment with \num{203} X (formerly Twitter) users on participants' self-reported willingness to like different kinds of content. We find no detectable platform-level increase in likes for high-reputational-risk content (Study 1). Additionally, while participants in the survey experiment, particularly those with higher education and income, report modest increases in willingness to like high-reputational-risk content under private versus public visibility, these increases do not lead to significant changes in the group-level average likelihood of liking posts (Study 2). Taken together, our results suggest that hiding liker identity produces a limited behavioral response at the platform level, which may be caused by a gap between user intention and behavior.
\end{abstract}


\maketitle

\section{Introduction}
\label{sec:intro}
Social media platforms scaffold public expression through interface affordances that make some actions easy, legible, and rewarding while rendering others costly or invisible~\cite{bucher2018affordances,ronzhyn2023defining,devito2017platforms}. 
Among these, the visibility of engagement signals is central: visible cues act as social proof and reputation broadcasts, shaping both audiences' inferences and authors' self-presentation strategies~\cite{marder2016like}.

In June 2024, \X implemented a major design change by making liker identities private: users can still like posts, and authors can see who liked them, but third parties can no longer inspect others' likes (see illustration in \Cref{sec:research_overview}).
This policy directly reconfigures the platform's visibility affordance for liking without altering the act of posting, offering a rare, platform-scale case to study how privacy of engagement changes behavior.

But why might hiding liker identities matter?
First, reputational and normative pressures can deter the visible endorsement of sensitive content, leading to self-censorship~\cite{das2013self,devito2017platforms}. 
Second, visible engagement metrics amplify bandwagon dynamics and moralized responses~\cite{chuai2025community,chuai2024did}. 
If reputational risk is a primary brake on liking, privatizing the liker's identity should \emph{increase} engagement with high-reputational-risk content (content that users would feel social pressure not to engage in public; \eg, partisan or lewd posts), while leaving low-reputational-risk content largely unchanged.

Beyond social dynamics, like visibility also matters for the platform's algorithmic layer. 
Engagement signals such as likes, reposts, and replies are integral inputs to ranking and recommendation systems that determine which posts are surfaced to users~\cite{tenenboim2022comments,gerlitz2013like}. 
Publicly visible likes historically functioned as both a \emph{social} and an \emph{algorithmic} feedback loop: they indicated popularity to users while signaling content quality to the platform. By privatizing liker identities, \X decouples these two layers, reducing the social signaling function while leaving the algorithmic weight of likes intact. This shift could alter how engagement propagates through the feed: if users become marginally more willing to like controversial content, the ranking system may amplify it even without visible social proof. Conversely, if liking remains concentrated among a small subset of users or automated accounts, the algorithm's exposure dynamics may stay largely unchanged~\cite{ferrara2016rise,zhu2016attention}. Understanding whether this design change modifies not only users' visible behavior to controversial content but also the algorithmic distribution of content is key to assessing its civic and informational consequences.

Here, we use \X's policy change as a quasi-experiment and ask: \emph{did making likes private change the number of likes received by high-reputational-risk content relative to low-reputational-risk content?} We answer this question with an observational analysis of \num{153704} posts that are created by \num{1045} accounts and have received over 324 million likes spanning eight weeks pre-change and four weeks post-change. 
We estimate effects with Difference-in-Differences (DiD) approach across two complementary designs: (i) between groups (high- vs.\ low-reputational-risk accounts) and (ii) within posts (likes vs.\ reposts), and probe pre-trends and inference robustness with HonestDiD and equivalence test. Additionally, to address potential mechanism plausibility, we run a survey experiment in which participants rate their likelihood of liking posts on high-reputation-risk and low-reputational-risk topics under public vs.\ private visibility.

Across both DiD designs, weekly and aggregated effects are statistically indistinguishable from zero: the privacy change did not yield a detectable platform-level increase in likes to high-reputational-risk content, nor a shift in likes relative to reposts. In the survey, participants, especially well-educated and high-income ones, report modest within-person increases in willingness to like controversial content under private visibility, but group means remain similar across visibility conditions. 
Together, our results suggest a limited behavioral response at scale despite stated preferences, which is consistent with an intention–behavior gap 
% and the concentration of engagement among specialized, high-usage, or automated accounts.

Our contributions are three-fold: (i) A causal estimate of how privatizing liker identities affects engagement with high-reputational-risk content, using two DiD strategies with robustness checks. (ii) A mixed-methods triangulation that contrasts platform-level behavior with user-level stated preferences to illuminate mechanisms. (iii) Design implications for visibility interventions: hiding \emph{who} liked may be insufficient to meaningfully shift sensitive engagement when other incentive and distributional forces dominate.
We discuss implications for moderation on social media platforms: visibility changes that target reputational risk may require complementary levers (\eg, distribution, feedback timing, or audience segmentation) to move behavior in consequential ways.

\section{Related Work}
\subsection{Social media affordances}
Social media affordances refer to the perceived, actual, and imagined properties of platforms that emerge through the interplay of technological, social, and contextual factors, shaping and constraining how users engage with them~\cite{ronzhyn2023defining,bucher2018affordances}. These affordances are generally conceptualized along two levels: high-level affordances capturing  the abstract capabilities of platforms, and low-level affordances, embedded in the material design of specific features, such as buttons, screens, and interactive elements~\cite{bucher2018affordances,treem2013social}.

The four-item high-level taxonomy of affordances -- visibility, editability, persistence, association -- applies to a range of social media platforms and covers the most significant and controversial aspects of the use of social media~\cite{ronzhyn2023defining}. Visibility refers to the extent to which information, behaviors, and connections are observable to others within the platform. Editability captures users' ability to craft, revise, and refine content before or after publication, allowing for greater control over self-presentation. Persistence denotes the durability of digital traces over time, as content remains accessible and searchable long after initial creation. Finally, association describes the capacity of social media to link individuals, content, and communities through features such as tagging, following, or connecting. To better understand self-presentation on social media platforms, prior research proposed an affordance-based framework, which includes three main components related to the self (presentation flexibility, content persistence, identity persistence), other actors (content association, feedback directness), and the audience (transparency and visibility control)~\cite{devito2017platforms}.

The affordances of social media platforms are constrained by engagement features, \ie, low-level affordances~\cite{bucher2018affordances}. Likes, comments, and shares are three main engagement features across many social media platforms such as \X (formerly Twitter), Facebook, and TikTok. The value affordances of these engagement features are summarized by previous research, and each feature is associated with specific trade-offs~\cite{scharlach2023value}. Specifically, the Like feature encourages recognition and positivity but undermines authenticity; the Comment/Reply feature fosters feedback and diversity but invites conflict; and the Sharing feature strengthens togetherness but can pressure users and threaten privacy. For instance, by rewarding specific types of posts through ``likes,'' users not only encourage others to produce similar content but also shape their own content exposure, as platform algorithms curate their feeds based on prior engagement patterns~\cite{tenenboim2022comments,gerlitz2013like}.

\subsection{Content moderation and self-censorship}
Social media platforms provide users with considerable flexibility for expression, yet they enable the creation and dissemination of controversial or harmful content, such as misinformation, hate speech, and harassment~\cite{menczer2016spread,aimeur2023fake,vosoughi2018spread,matamoros2021racism,mathew2019spread,maarouf2024virality,wong2021standing}. Consequently, platforms face growing pressure to design effective moderation mechanisms that mitigate harm while safeguarding freedom of expression. Existing content moderation interventions are largely designed to act upon or reshape the affordances of social media. For instance, the use of warning labels, annotations, downranking, and visibility reduction directly targets the visibility affordance by limiting the exposure of problematic or misleading content~\cite{chuai2024did,chuai2024community,martel2023misinformation,epstein2020will,gillespie2022not,horta2023automated,zhang2025commenotes}. Similarly, the editing or removal of posts and the deplatforming of user accounts are tied to the editability and persistence affordances, as they determine whether and how content can be altered, deleted, or remain accessible over time~\cite{beadle2025edit,chandrasekharan2022quarantined,horta2023deplatforming,horta2021platform,jhaver2019did,jhaver2021evaluating,mitts2022removal,ali2021understanding,chandrasekharan2017you,jhaver2019does}.

Notably, from the user perspective, visibility is a consequential affordance, constituting the social dimension of online interaction. High visibility can promote exposure to a broad audience but also intensifies social evaluation, leading users to self-censor content they perceive as controversial or reputationally risky~\cite{ronzhyn2023defining,devito2017platforms}. \emph{Self-censorship refers to the deliberate act of withholding or altering expression to avoid potential backlash or sanctions~\cite{das2013self}.} This behavior is reinforced by platform feedback systems and community-driven norms, such as post guidance~\cite{horta2025post}, community notes that flag misleading posts~\cite{chuai2024did,chuai2024community,zhang2025commenotes}, and collective reactions of moral outrage~\cite{chuai2025community}. For example, the display of community notes can increase the odds that authors delete their misleading posts by 103.4\%~\cite{chuai2024community}, demonstrating how visibility-based interventions can shape user behavior. 

Users' perceptions of being observed also influence their willingness to disclose personal/sensitive information. Heightened surveillance awareness amplifies privacy concerns and reduces the perceived value of sharing, prompting protective behaviors such as self-censorship~\cite{kim2025watching,saha2024observer}. Additionally, the perceived platform audience composition plays a crucial role in disclosure decisions involving sensitive content~\cite{birnholtz2020sensitive}. On \X, users strategically navigate visibility boundaries by toggling between public and protected account settings to manage audience exposure and personal disclosure~\cite{kekulluoglu2022understanding,kekulluoglu2023twitter}.

From the audience or observer perspective, anonymous sharing is a popular choice, especially for controversial content, suggesting that users rely on anonymity to mitigate reputational risks~\cite{zhang2014anonymity}. Social media anonymity can also enhance moral courage, as it reduces the perceived risk of adverse consequences for taking moral actions~\cite{pan2023we}. Beyond anonymity, the visibility of engagement metrics significantly influences how users interact with content. For example, the public visibility of likes makes this form of engagement more susceptible to normative social influence than less visible behaviors, such as clicking or viewing~\cite{huang2025more,lee2016makes}. Users also strategically manage their public personas through visible engagement cues, for instance, withholding likes on politically sensitive content to avoid potential reputational consequences~\cite{marder2016like}.

\textbf{Research gap our study aims to address:}
On \X, recent platform change -- specifically, the shift from public to private likes -- represents a significant reconfiguration of the visibility affordance~\cite{wsp2024private,sleeper2013post}: users can see posts they have liked (but others cannot); like count and other metrics for users' own posts will still show up under notifications; users no longer see who liked someone else's post; and post authors can see who liked their posts. This design alteration has the potential to reshape patterns of engagement by reducing the normative and reputational pressures associated with visible liking behavior to the controversial content. Examining how user interactions change following this policy shift is therefore crucial for understanding how visibility mediates expression and self-censorship in contemporary social media environments, which presents the aim of this study.

\begin{figure*}
  \includegraphics[width=\textwidth]{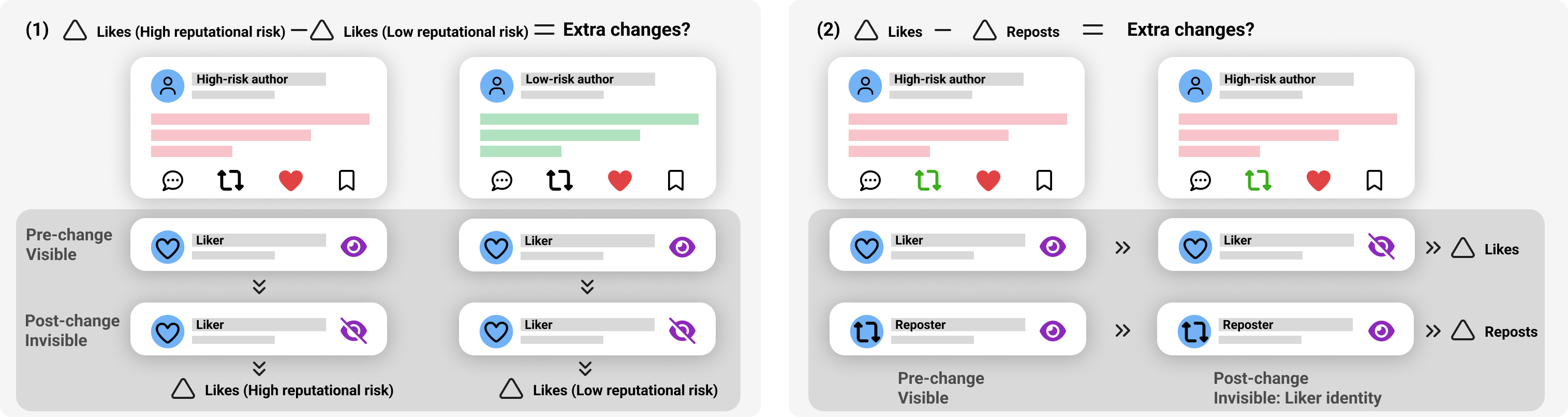}
  \caption{Research overview. \X made public likes private on June 12, 2024. After the change, the identities of likers are visible only to themselves and the post authors. To evaluate the behavioral impact of this change, we estimate (1) the additional changes in likes received by high-reputational-risk accounts after the design change, relative to the pre-change period and compared to low-reputational-risk accounts (between-group); and (2) the additional changes in likes to high-reputational-risk accounts after the design change, relative to the pre-change period and compared to reposts on the same accounts (within-group).}
  \label{fig:visibility_change}
\end{figure*}

\section{Materials and Methods}
\label{sec:methods}

\section{Research Overview}
\label{sec:research_overview}

After the likes made private, the identities of likers are visible only to themselves and the post authors, while the identities of reposters keep publicly visible. We use \cref{fig:visibility_change} to explain how we estimate the effect of the visibility change on users' liking behavior by comparing (i) high-reputational-risk accounts vs. low-reputational-risk accounts and (ii) likes vs. reposts. The descriptions for our observational data is reported in \Cref{tab:data_summary}.

\begin{table*}
\centering
\setlength{\tabcolsep}{5pt}
\caption{Overview of accounts, posts, and their associated engagement of likes and reposts.}
\label{tab:data_summary}
\begin{tabular}{l*{4}{c}}
\toprule
&{(1)} & {(2)} & {(3)} & {(4)}\\
&{\#Accounts}&{\#Posts}&{\#Likes}&{\#Reposts}\\
\midrule
\multicolumn{5}{l}{\underline{High-reputational-risk}}\\
\quad Extremist & {30} & {8,520} & {20,314,165} & {4,591,610}\\
\quad OnlyFans & {138} & {34,912} & {16,430,334} & {1,636,391}\\
\quad Political influencer & {66} & {30,444} & {187,296,366} & {33,332,385}\\
\quad Politician & {562} & {66,408} & {44,847,331} & {9,961,371}\\
\multicolumn{5}{l}{\underline{Low-reputational-risk}}\\
\quad Actor & {48} & {1,323} & {9,289,093} & {1,069,448}\\
\quad Gamer & {72} & {2,749} & {7,426,351} & {245,959}\\
\quad Singer & {65} & {1,940} & {34,209,734} & {4,959,676}\\
\quad Tech influencer & {64} & {7,408} & {4,238,341} & {224,299}\\
{\underline{Total}} & {1,045} & {153,704} & {324,051,715} & {56,021,139}\\
\bottomrule
\end{tabular}
\end{table*}

\subsection{Observational data collection}
\label{sec:data_collection}
We consider \X's accounts belonging to categories that may or may not be affected by changes in self-censoring behavior following \X's feature change, and classify them into high-reputational-risk and low-reputational-risk groups~\cite{zhang2014anonymity,elmimouni2025exploring}. This distinction is not intended to judge the nature of the content itself, but rather to capture that users may hesitate to publicly engage with certain accounts due to perceived social or reputational pressures (\eg, interactions with politically sensitive accounts). The high-reputational risk accounts include:
\begin{itemize}[leftmargin=*]
    \item SPLC extremist accounts ($N=44$). Accounts belonging to individuals the Southern Poverty Law Center (SPLC) listed as extremists~\cite{splc2025extremists}. The list includes 144 extremists. We manually search for their accounts on \X and successfully find 44 active ones throughout the observation period.
    \item Politicians ($N=632$). We obtain a dataset of politicians' \X handles from Wikidata~\cite{wiki2025query}. Specifically, we consider former U.S. Presidents as well as members of the 118th U.S. Congress (both Senate and House of Representatives). Additionally, we supplement the political ideology scores for the collected politicians~\cite{lewis2025voteview}.
    \item Political influencers on \X ($N=69$). AllSides X Influencer Bias Chart (see \cref{fig:influencer_bias}, \Cref{supp:political_influencer}) includes bias ratings for the most influential content creators on \X~\cite{allsides2025x}. Based on this chart, we collect their linked account handles on \X.
    \item OnlyFans accounts ($N=229$). We consider accounts associated with sex workers on OnlyFans. The official OnlyFans handle regularly promotes its member accounts. Given this, we first collect all the posts created by the OnlyFans official account during 2023 and extract the member accounts mentioned in the posts.
\end{itemize}
In addition, we consider the following low-reputational-risk accounts:
\begin{itemize}[leftmargin=*]
    \item Tech influencers ($N=71$). We select tech influencers with active \X accounts based on a 2025 list of top tech influencers~\cite{feedspot2025top}.
    \item Gamers ($N=100$) / Singers ($N=100$) / Actors ($N=100$). We again query Wikidata to search for gamers, singers, and actors. Subsequently, we select the top 100 professionals with the highest number of followers on \X for each of the three categories.
\end{itemize}

We use the \X Pro API's full-archive search endpoint to collect posts created by the selected accounts between eight weeks before and four weeks after likes were made private. As a result, we successfully collect \num{153704} posts across \num{1045} accounts (see details in \Cref{tab:data_summary}). We extract accounts' build-in features and posting preferences based on the collected posts during eight weeks of pre-change period for account heterogeneity analysis. The accounts' build-in features include the number of followers and their account age. We also consider accounts' posting frequency (\ie, the number of posts per account during pre-change period), user likes (the average like count per account during pre-change period). Additionally, given that sentiments are an important driver for online engagement~\cite{chuai2022anger}, we emply a state-of-the-art sentiment classification model to calculate positive and negative sentiments for the collected posts~\cite{camacho2022tweetnlp}. Then, we measure accounts' preference on posting positive or negative posts based on the average positive or negative sentiments in their posts during pre-change period.

\subsection{DiD specification}
\label{sec:did_model}

We employ DiD design from two complementary perspectives: 
(i) \textit{Between-group comparison.} 
We compare liking patterns between high-reputational-risk and low-reputational-risk accounts to test whether the visibility change disproportionately affected engagement with socially sensitive posts. 
We assume that the policy's impact should be larger for high-reputational-risk accounts, those for which public engagement carries greater reputational risks, making them our treatment group. Low-reputational-risk accounts, for which engagement is less socially costly, serve as the control group. 
We then estimate the differential change in the number of likes these two groups received before and after the policy shift.
(ii) \textit{Within-group comparison across engagement types.} 
We further compare liking behavior with other forms of engagement that were not directly affected by the policy change. Given that likes and reposts have shared motivations and affordances~\cite{scharlach2023value} and that likes are strongly correlated with reposts in our dataset ($r=0.853$, $p<0.001$; 95\% CI: [\num{0.851}, \num{0.854}]), we restrict our analysis to high-reputational-risk accounts and examine whether likes increased relative to reposts after the policy shift, isolating the effects specific to private visibility from broader engagement dynamics (see the research overview in \cref{fig:visibility_change}).

The parallel trends is a key assumption for DiD identification: in the absence of the policy intervention, the difference between the treatment and control outcomes would have evolved similarly over time. In our first comparison, this implies that the gap in likes between high- and low-reputational-risk accounts would have remained stable absent the visibility change; in the second, that likes and reposts would have evolved in parallel for high-reputational-risk accounts. 
Together, these complementary designs allow us to attribute observed shifts in liking behavior to the introduction of private like visibility rather than to general engagement trends on the platform.

Our DiD approach includes both weekly leads-and-lags and aggregated pre-post specifications. The leads-and-lags model estimates the weekly dynamics of engagement before and after the policy change, allowing us to assess the parallel trends assumption and track the temporal evolution of treatment effects. The aggregated pre-post model captures the overall effect of making likes private on \X. High-reputational-risk posts or likes serve as the treatment group, with low-reputational-risk posts or reposts acting as controls. Additionally, post-specific and week-specific fixed effects are included in both models to account for unobserved heterogeneity across posts and time. Given that likes and reposts are count variables, we estimate our models using negative binomial regression. 

\textbf{Leads-and-lags specification.} 
To examine temporal dynamics before and after likes were made private, we estimate a leads-and-lags Difference-in-Differences (DiD) specification:
\begin{equation}
\begin{aligned}
    \var{log(E(Y_{i,t})|\bm{x}_{i,t})} = & \beta_{0} + \beta_{1}D_{i} + \bm{\beta}_{2}'\textbf{Before}_{t} + \bm{\beta}_{3}'\textbf{After}_{t} + \bm{\beta}_{4}'(D_{i} \times \textbf{Before}_{t}) + \bm{\beta}_{5}'(D_{i} \times \textbf{After}_{t}) + \alpha_{i},
\end{aligned}
\end{equation}
where $D_{i}$ is a dummy variable indicating whether post $i$ belongs to the high-reputational-risk group ($=1$) or the low-reputational-risk group ($=0$). 
When comparing likes with reposts within the same posts, $D_{i}$ instead indicates whether the engagement type is a like ($=1$) or a repost ($=0$).

The vectors $\textbf{Before}_{t}$ and $\textbf{After}_{t}$ denote week dummies relative to the implementation date of private likes. 
Specifically, $\textbf{Before}_{t}$ covers weeks $-8$ to $-2$, with week $-1$ serving as the reference period, while $\textbf{After}_{t}$ spans weeks $+1$ to $+4$. 
Thus, $\bm{\beta}_{2}$ and $\bm{\beta}_{3}$ capture time fixed effects, and $\alpha_{i}$ denotes post-specific fixed effects. 
Finally, the interaction coefficients $\bm{\beta}_{4}$ represent pre-treatment (lead) effects, used to assess the parallel trends assumption, whereas $\bm{\beta}_{5}$ capture post-treatment (lag) effects, measuring the weekly impact of the private-like visibility policy on liking behavior.

\textbf{Aggregated pre-post specification.} 
We examine the aggregated effect of the private visibility based on the following DiD specification with two periods, \ie, pre- and post-policy change:
\begin{equation}
\begin{aligned}
    \var{log(E(Y_{i,t})|\bm{x}_{i,t})} = \beta_{0} + \beta_{1}D_{i} + \beta_{2}\text{After}_{t} + \beta_{3}D_{i} \times \text{After}_{t} + \alpha_{i} + \gamma_{t},
\end{aligned}
\end{equation}
where $\text{After}_{t}$ is a post-treatment indicator that equals to 1 after the implementation of private visibility and 0 before. The coefficient estimate of $\beta_{3}$ captures the aggregated effect of the private visibility on the liking behavior. $\alpha_{i}$ and $\gamma_{t}$ represent post-specific and weekly-specific fixed effects, respectively.

\textbf{Transformation of effect size.}
The coefficients estimated for the DiD terms in the two-period and multi-period regression models are the natural logarithms of the ratios for the number of likes with the treatment of private visibility compared to the number of likes that are expected to receive with public visibility during the post-treatment period. We exponentially transform the coefficient estimates of the DiD terms in the models and measure the effect size of the private visibility as:
\begin{equation}
\begin{aligned}
    \text{ATT} = e^{\beta} - 1,
\end{aligned}
\end{equation}
where $\beta$ is the coefficient estimate for the specific DiD term. ATT indicates the ratio of extra change of the number of likes to posts with the treatment of private visibility relative to the number of likes that are expected to receive with public visibility. We used this indicator to analyze the effect of private visibility on liking behavior.

\subsection{User experiment}
\label{sec:exp_setup}

We design a within-subject survey experiment in which each participant was shown with topics that varied in reputational risk -- that is, the potential reputational cost of publicly engaging with them. 
For each post, participants rated how likely they would be to ``like'' it under two conditions: (i) when likes were publicly visible and (ii) when likes were private (visible only to the author and the liker).
This design allows us to compare self-reported willingness to engage across visibility settings and exposure levels within the same individuals.  

The set of topics is selected to capture both high- and low-reputational-risk contexts. Controversial, high-reputational-risk topics include abortion, the Black Lives Matter (BLM) movement, immigration and border security, transgender rights, and gun control.  
Additionally, non-controversial, low-reputational-risk topics include art, charities working to end hunger, and pop culture.\footnote{we also select accounts to conduct account-based analysis according to the categories listed in Study 1. The results remain robust and consistent with our main findings.}  

We recruit $N=203$ participants through the Prolific platform, restricting the sample to \US-based users who reported being active \X users. 
To approximate the demographic composition of the \X user base in the \US, we employ a stratified sampling procedure based on age, gender, education, political leaning, living area, and household income, using population estimates reported in the 2025 Pew Research Center survey on social media use~\cite{pew2024socialmedia}. While exact stratification is not possible due to platform-level constraints, we subsequently apply iterative proportional fitting algorithm, implemented with ipfn package in Python, to reweight the sample to match Pew's demographic margins (see \Cref{supp:reweighting} for details).

\noindent Additionally, we conduct subgroup analysis and examine the potential heterogeneity in participants' self-rated likelihood of liking high-reputational-risk posts and  based on their demographics. Specifically, we consider the following comparisons:
\begin{itemize}[leftmargin=*]
    \item \textit{Age}. Younger participants: aged 18--29; older participants: aged 30 or above.
    \item \textit{Gender}. Male vs. female participants.
    \item \textit{Education}. lower-educated participants: with a high school education or less; higher-educated participants: with a college education or higher.
    \item \textit{Political leaning}. Left-leaning vs. right-leaning participants.
    \item \textit{Living area}. Urban vs. non-urban participants.
    \item \textit{Household income}. lower-household-income participants: annual household income below \$70,000; higher-household-income participants: annual household income of \$70,000 or above.
\end{itemize}

\noindent All procedures were reviewed and approved by the authors' institutional ethics board.

\section{Results}
\subsection{Study 1: Observational analysis}
\X made ``likes'' private on June 12, 2024.
In Study 1, we use the \X Pro API's full-archive search endpoint to collect posts created by the high-reputational-risk (SPLC extremists, OnlyFans creators, political influencers, and politicians) and low-reputational-risk accounts (actors, gamers, singers, and tech influencers) between eight ($-8$) weeks before and four ($+4$) weeks after likes were made private (see details in \Cref{sec:methods} and \Cref{supp:methods}). As a result, we successfully collect \num{153704} posts across \num{1045} accounts (see summary statistics in \Cref{tab:data_summary}). Subsequently, we employ DiD approach to estimate the effect of private likes on users' liking behavior toward posts from high-reputational-risk accounts. To interpret the results, we exponentiate the DiD estimates to obtain the relative change in likes attributable to the policy change, \ie, average treatment effect on the treated (ATT; see model details in \Cref{sec:did_model}).

\begin{figure}
    \centering
    \captionsetup[subfloat]{font={bf, small}, skip=0pt, singlelinecheck=false, labelformat=simple, position=top}
    \subfloat[]{\includegraphics[width = .48\textwidth]{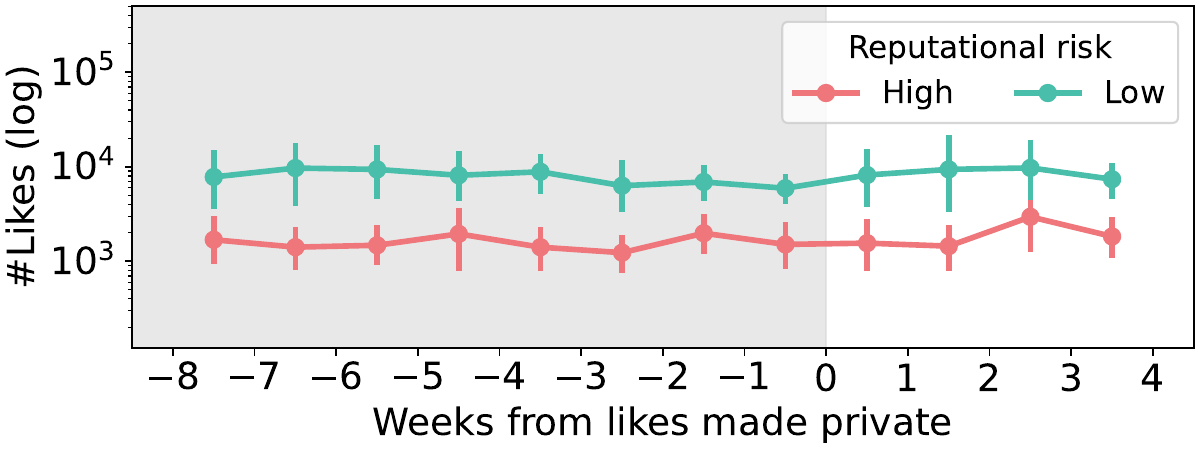}}
    \hfill
    \subfloat[]{\includegraphics[width = .48\textwidth]{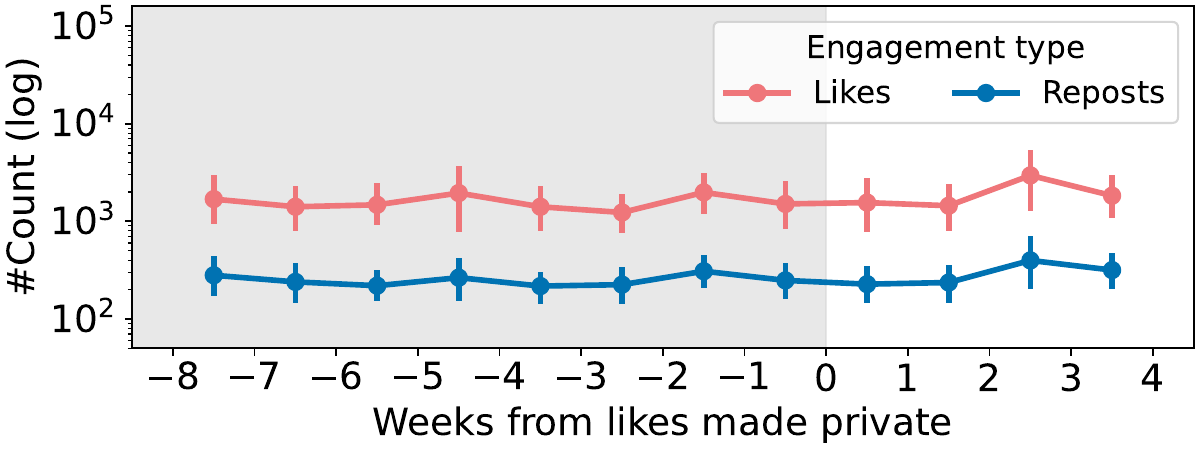}}
    
    \subfloat[]{\includegraphics[width = .48\textwidth]{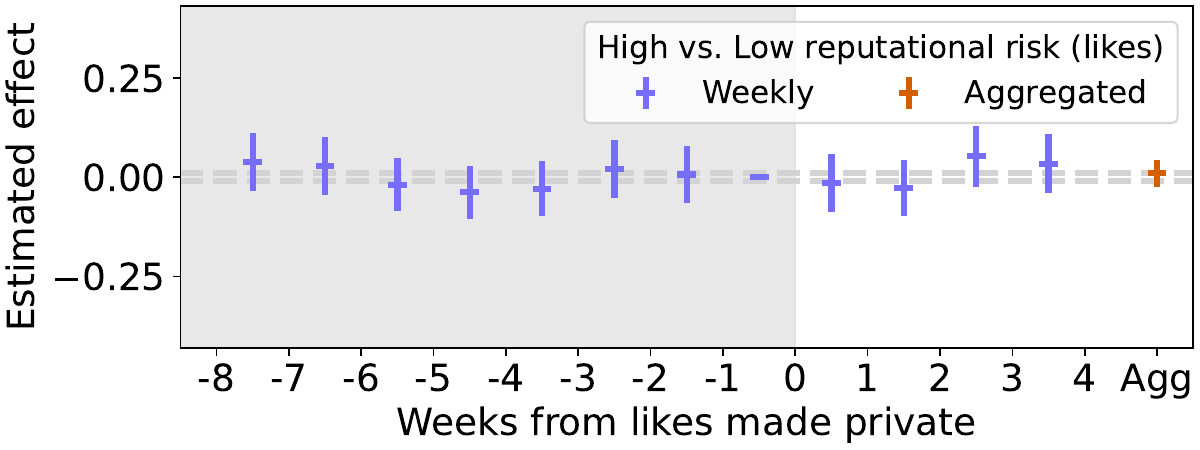}}
    \hfill
    \subfloat[]{\includegraphics[width = .48\textwidth]{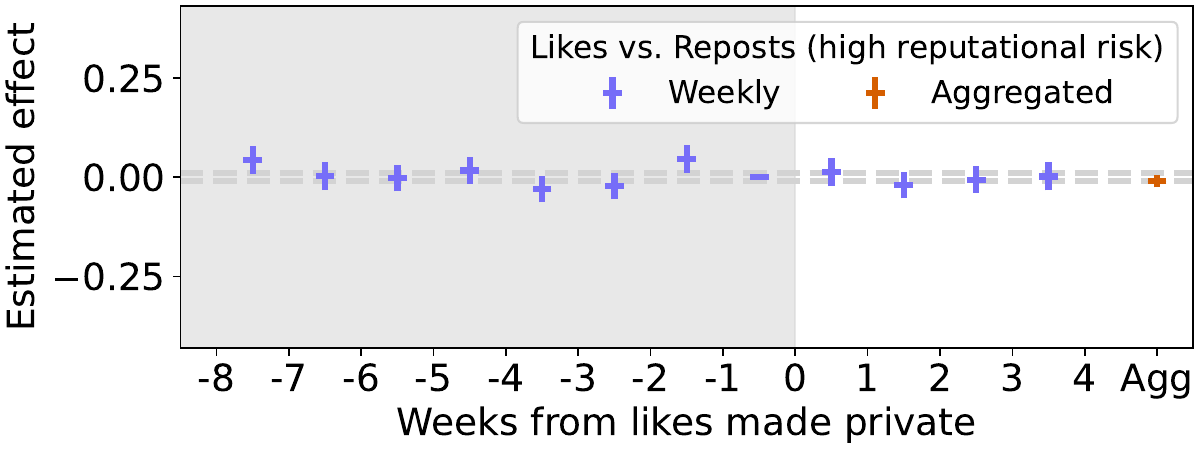}}
    \caption{Analysis of weekly effects of making likes private on the number of likes. (a)~The weekly averages of likes received by posts from high-reputational-risk and low-reputational-risk accounts. (b)~The weekly averages of likes and reposts for posts from high-reputational-risk accounts. (c)~The estimated weekly (and aggregated) effects of private visibility on likes to posts from high-reputational-risk accounts relative to low-reputational-risk accounts. (d)~The estimated weekly (and aggregated) effects of private like visibility on the number of likes for posts from high-reputational-risk accounts relative to the number of reposts. The error bars represent 95\% CIs. The grey dash lines in (c) and (d) represent equivalence bounds.}
    \label{fig:att_estimation}
\end{figure}

We start at examining the weekly averages of likes on high-reputational-risk posts relative to (i) the weekly averages of likes on low-reputational-risk posts (\cref{fig:att_estimation}a) and (ii) the weekly averages of reposts on high-reputational-risk posts (\cref{fig:att_estimation}b). Across both comparison, we observe that weekly engagement levels remain relatively stable following the shift to private like visibility, with no clear divergence between high-reputational-risk vs. low-reputational-risk groups or likes vs. reposts. These descriptive patterns provide an initial indication that making likes private did not trigger a large-scale shift in how users engage with high-reputational-risk content.

\subsubsection*{Weekly and aggregated DiD estimates}
\cref{fig:att_estimation}c reports the estimated weekly and aggregated effects of private visibility on the number of likes received by high-reputational-risk posts relative to low-reputational-risk posts. We find that all weekly effect estimates are close to zero (each $p > 0.05$), indicating no statistically significant change in likes to high-reputational-risk content following the policy change. The aggregated DiD estimate (ATT $=0.01$, $z=0.592$, $p=0.554$; 95\%CI: $[-0.023, 0.044]$) likewise shows a statistically non-significant effect and an effect size close to zero, implying that the overall effect of making likes private is negligible in both statistical and practical terms.
Similarly, we estimate the weekly and aggregated effects of private visibility on the number of likes relative to reposts for high-reputational-risk posts (\cref{fig:att_estimation}d). We find that the weekly effect estimates are not significantly different from zero, although we observe minor pre-trends exist in week $-8$ (ATT $=0.044$, $z=2.491$, $p=0.013$; 95\%CI: $[0.009, 0.08]$) and week $-2$ (ATT $=0.046$, $z=2.67$, $p=0.008$; 95\%CI: $[0.012, 0.082]$). The aggregated effect estimate (ATT $=-0.01$, $z=-1.359$, $p=0.174$; 95\%CI: $[-0.025, 0.005]$) is likewise small and not statistically significant, consistent with the null results in the comparison of likes between high-reputational-risk and low-reputational-risk posts.

\subsubsection*{Robustness checks}
To assess the robustness of our weekly DiD estimates, we (i) conduct an equivalence test to determine whether the non–statistically significant effects can be dismissed as practically null, and (ii) account for potential violations of the parallel trends assumption by applying the HonestDiD framework to obtain robust confidence intervals for the original estimates~\cite{chuai2025community,robertson2023negativity}. For the equivalence test, we set an effect-size bound interval of [$-0.01$, $0.01$], which reflects the smallest change in weekly liking behavior that we consider meaningful in this context. All weekly estimates fall within the bound interval (each $p > 0.05$; see \cref{fig:att_estimation}c), implying a true null effect.

Additionally, HonestDiD provides a framework for robust inference in DiD settings where the strict parallel trends assumption may not fully hold~\cite{chuai2024community,rambachan2019honest}. Rather than invalidating results when minor pre-trends are detected, HonestDiD allows for small deviations between treatment and control groups and adjusts the inference accordingly. Specifically, we implement the Conditional Least Favorable Hybrid (C-LF) method, assuming that any deviations from parallel trends in the post-treatment weeks do not exceed $\overline{M}$ times the largest deviation observed in the pre-treatment weeks. The results indicate that the robust estimates remain statistically indistinguishable from zero, reinforcing our finding that the private visibility did not significantly affect users' liking behavior toward high-reputational-risk content (see \cref{fig:honest_did} in \Cref{supp:did_robustness} for details). 

Furthermore, we conduct additional robustness checks in terms of changes relative to accounts' like count baselines and shifted post-treatment period, the results remain consistent with our findings (see details in \Cref{supp:did_robustness}).

\begin{figure}
    \centering
    \captionsetup[subfloat]{font={bf, small}, skip=0pt, singlelinecheck=false, labelformat=simple, position=top}
    \subfloat[]{\includegraphics[width = .48\textwidth]{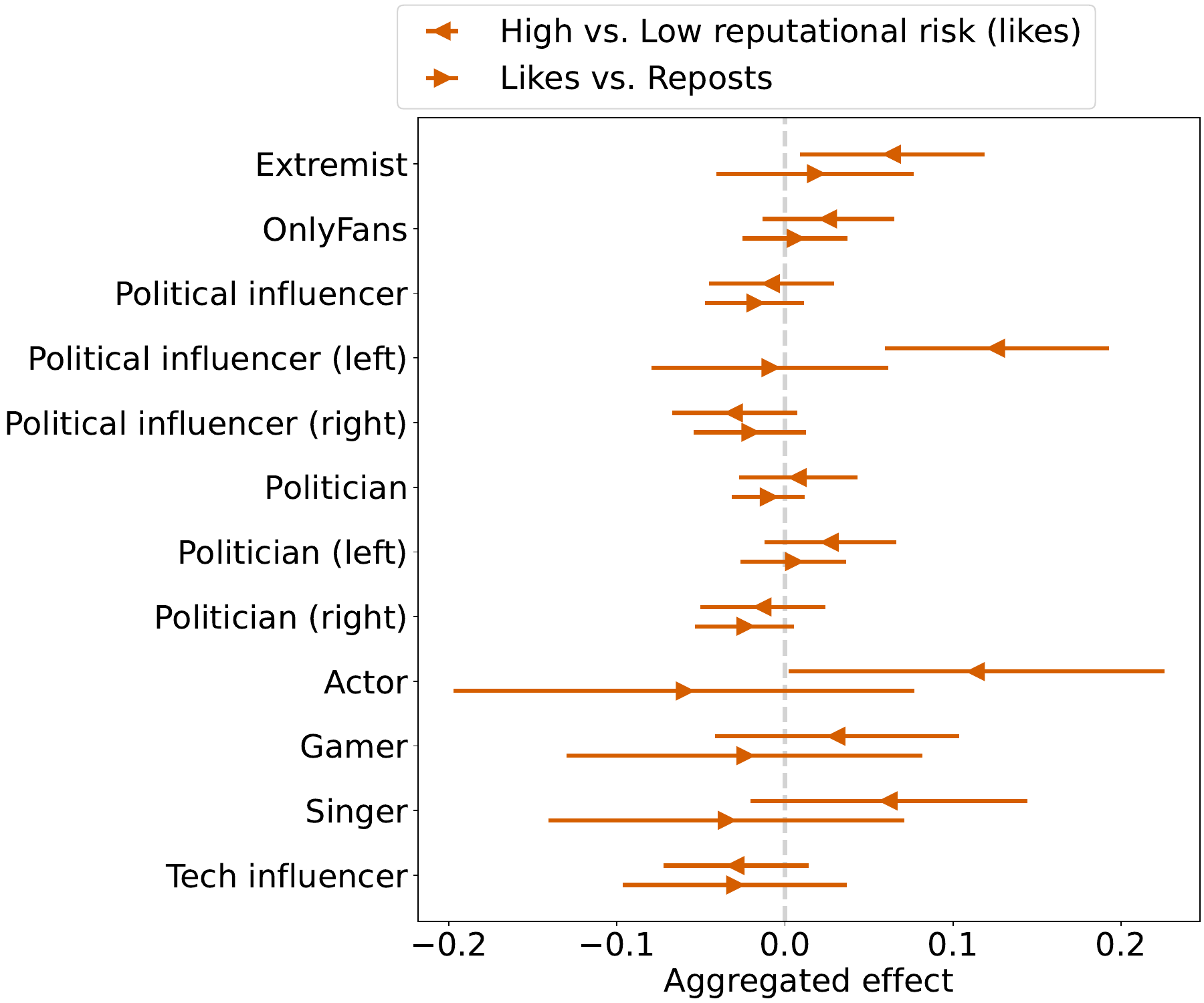}}
    \hfill
    \subfloat[]{\includegraphics[width = .48\textwidth]{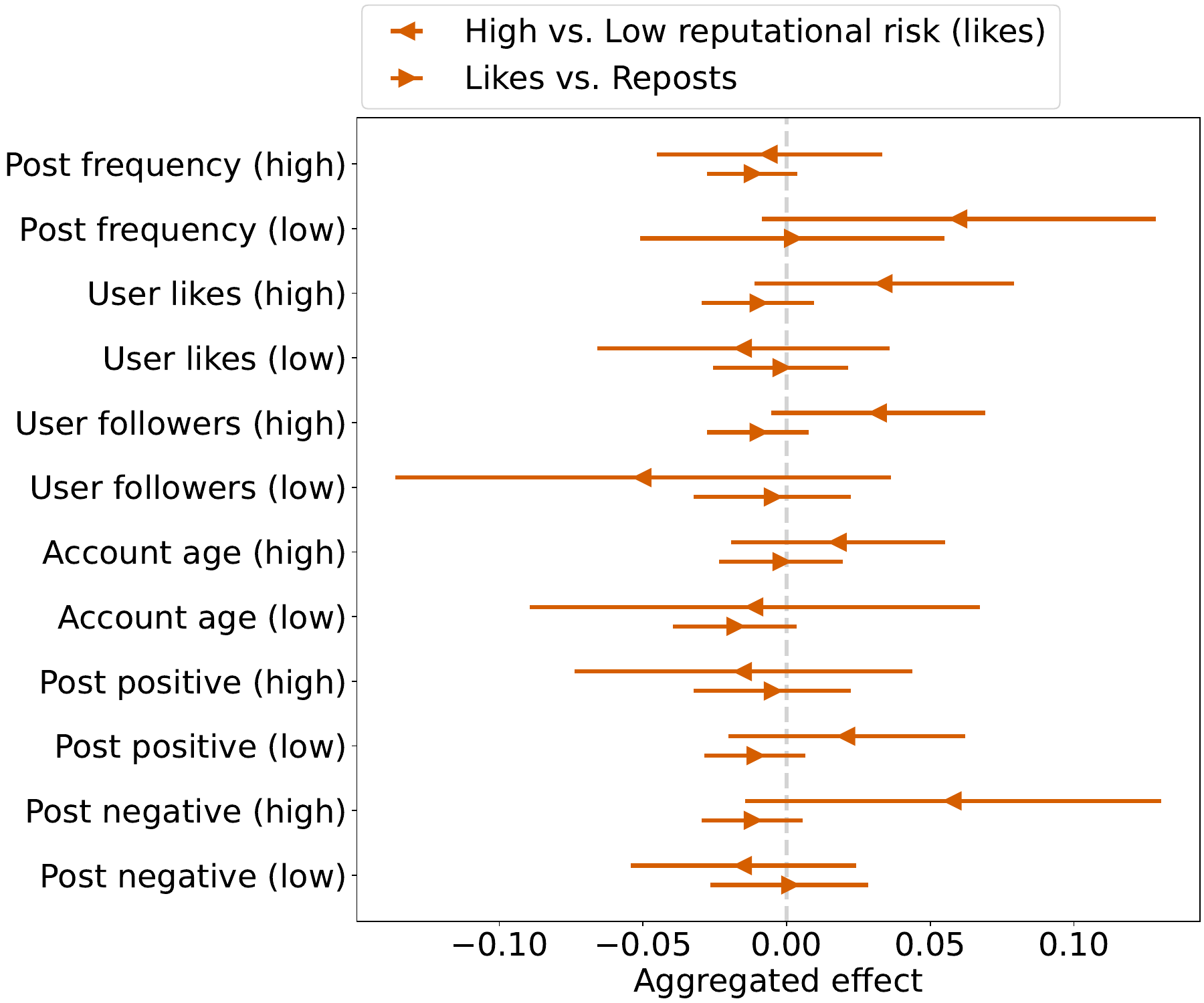}}
    \caption{Sensitivity analysis. (a)~The estimated aggregated effects across account categories in Study 1. (b)~The estimated aggregated effects across build-in account profiles and post features on \X. The error bars represent 95\% CIs.}
    \label{fig:agg_effects_sensitivity}
\end{figure}

\subsubsection*{Sensitivity analysis}
We conduct sensitivity analysis to examine how the aggregated effects of private visibility on the number of likes vary across account categories and engagement types (see \cref{fig:agg_effects_sensitivity}a). When comparing likes to high-reputational-risk posts vs. low-reputational-risk posts, we find that the estimated aggregated effects are significantly positive for SPLC extremists (ATT $=0.064$, $z=2.401$, $p=0.016$; 95\%CI: $[0.012, 0.12]$), left-leaning political influencers (ATT $=0.126$, $z=3.927$, $p<0.001$; 95\%CI: $[0.061, 0.195]$), and actors (ATT $=0.114$, $z=2.086$, $p=0.037$; 95\%CI: $[0.007, 0.232]$), while other aggregated DiD estimates are not statistically significant (each $p > 0.05$). However, importantly, when comparing likes with reposts for the same high-reputational-risk posts, the estimated aggregated effects are consistently statistically indistinguishable from zero (each $p > 0.05$). This suggests that the observed moderate effects in the high-reputational-risk vs. low-reputational-risk comparison for likes do not reflect a genuine response to the private visibility of likes, but instead stem from broader category-level differences unrelated to the policy change.

Additionally, we consider the accounts' heterogeneity based on the build-in features and their posting preferences. Specifically, the posting frequency (the number of posts), user likes (average like count per account), user followers, account age, posting positive (average positive sentiment in posts from an account), and posting negative (average negative sentiment in posts from an account) are included. All of these features are calculated based on the posts created during the pre-change period (see \Cref{sec:data_collection} for details). The aggregated ATT estimates across the account features are shown in \cref{fig:agg_effects_sensitivity}b. We find that all of the estimates are not significantly different from zero (each $p > 0.05$). This means that our findings are not subject to accounts' influence and their posting preferences on \X.

\subsection*{Study 2: User Study}
To validate our findings observed in Study 1, we further conduct a within-subject user study ($N=203$) to investigate whether the awareness of private visibility policy influences participants' liking behavior. Each participant was shown with topics that varied in reputational risk (\ie, the potential reputational cost of publicly engaging with them). Here, the high-reputational-risk topics include abortion, the Black Lives Matter (BLM) movement, immigration and border security, transgender rights, and gun control; the low-reputational-risk topics include art, charities working to end hunger, and pop culture (see details in \Cref{sec:exp_setup}). 

\begin{figure}
    \centering
    \includegraphics[width=.48\linewidth]{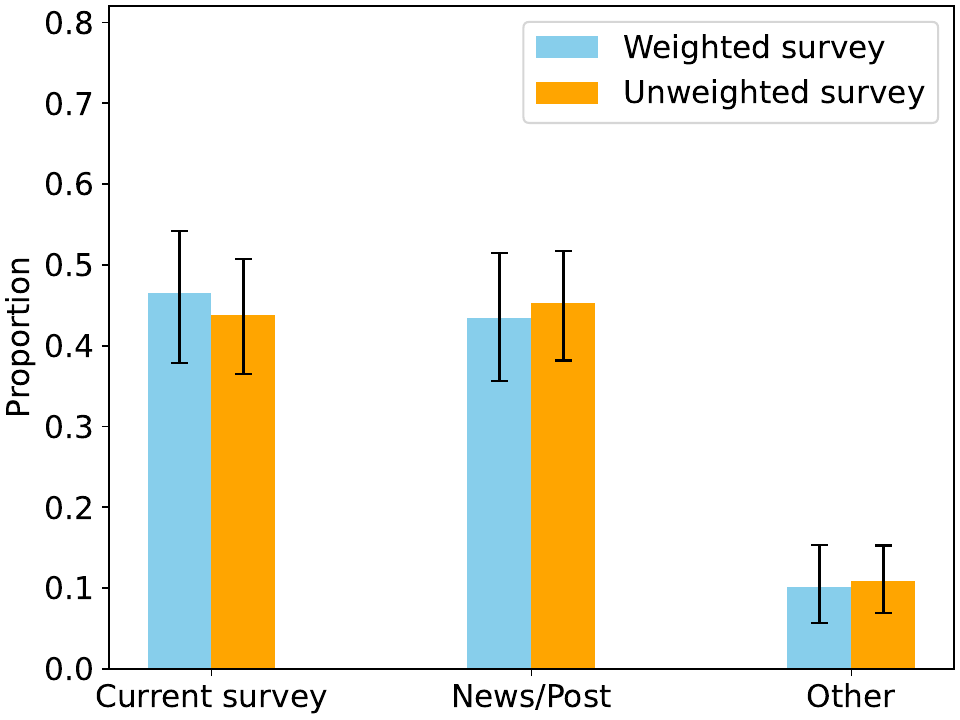}
    \caption{Reported channels through which participants first became aware of \X's change to private like visibility. Error bars represent 95\% CIs.}
    \label{fig:get_aware}
\end{figure}

We start by analyzing how participants first became aware of the private visibility policy on the liking behavior (\cref{fig:get_aware}). We find that approximately half of participants (mean $=\num{0.434}$, bootstrap=500; 95\% CI: [\num{0.357}, \num{0.515}]) reported learning about the change through news coverage or posts on \X, while a comparable portion (mean $=\num{0.465}$, bootstrap=500; 95\% CI: [\num{0.378}, \num{0.542}]) indicated that they first became aware of the policy change through our survey. This finding suggests that although the policy change received moderate public attention when the visibility became private, a substantial portion of users may not have been immediately informed or conscious of the modification in the visibility of their likes. In the following analysis, we analyze participants' self-rated likelihood of liking posts on both high-reputational-risk and low-reputational-risk topics. Additionally, we investigate whether participants who were previously aware of the policy change through news/posts (\ie, media-informed participants) and those who learned about it through other channels, \eg, our survey, behave differently to compare the long-term and short-term effects of the private visibility.

\begin{figure}
    \centering
    \captionsetup[subfloat]{font={bf, small}, skip=0pt, singlelinecheck=false, labelformat=simple, position=top}
    \subfloat[]{\includegraphics[width = .69\textwidth]{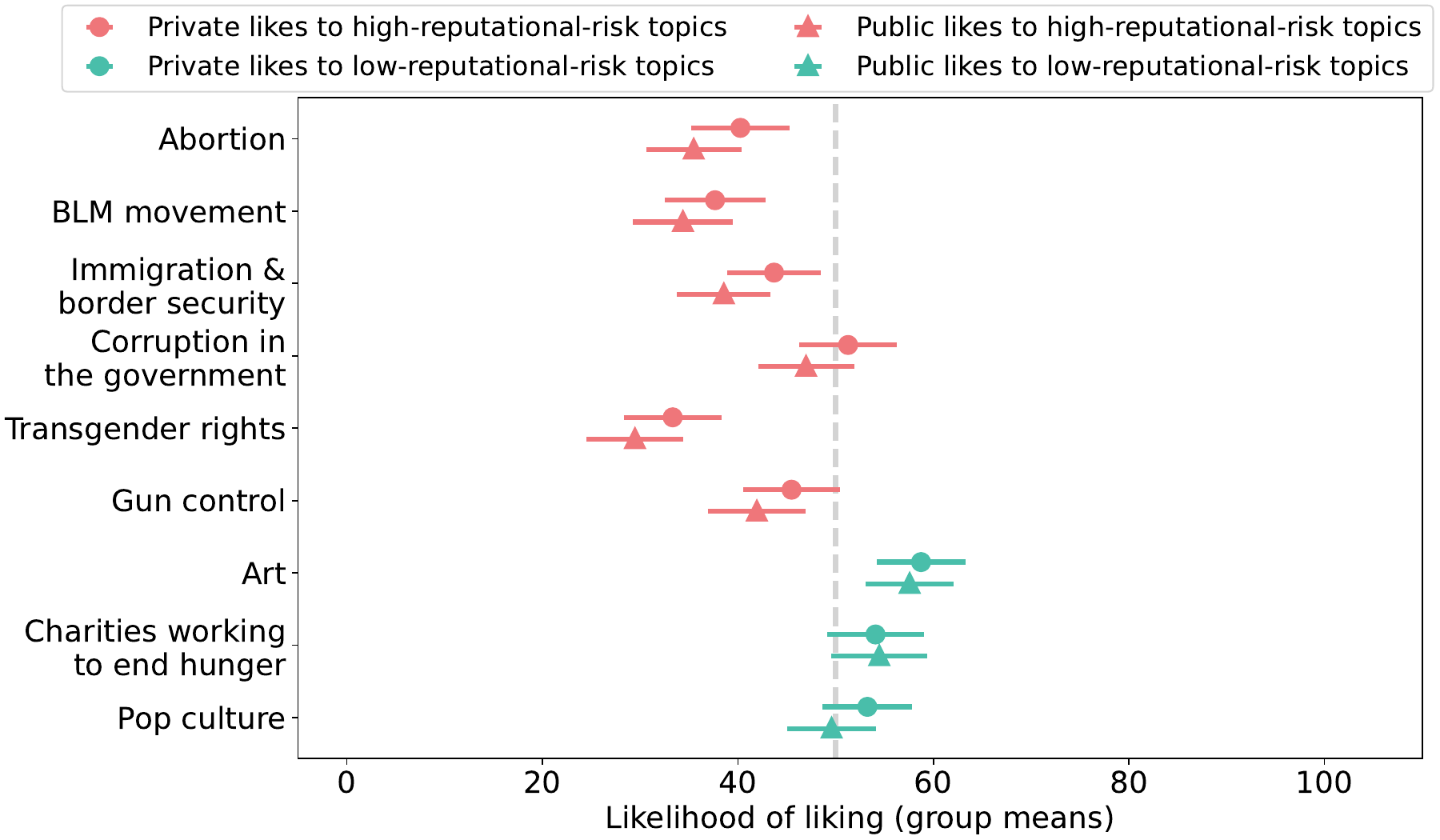}}
    \hfill
    \subfloat[]{\includegraphics[width = .288\textwidth]{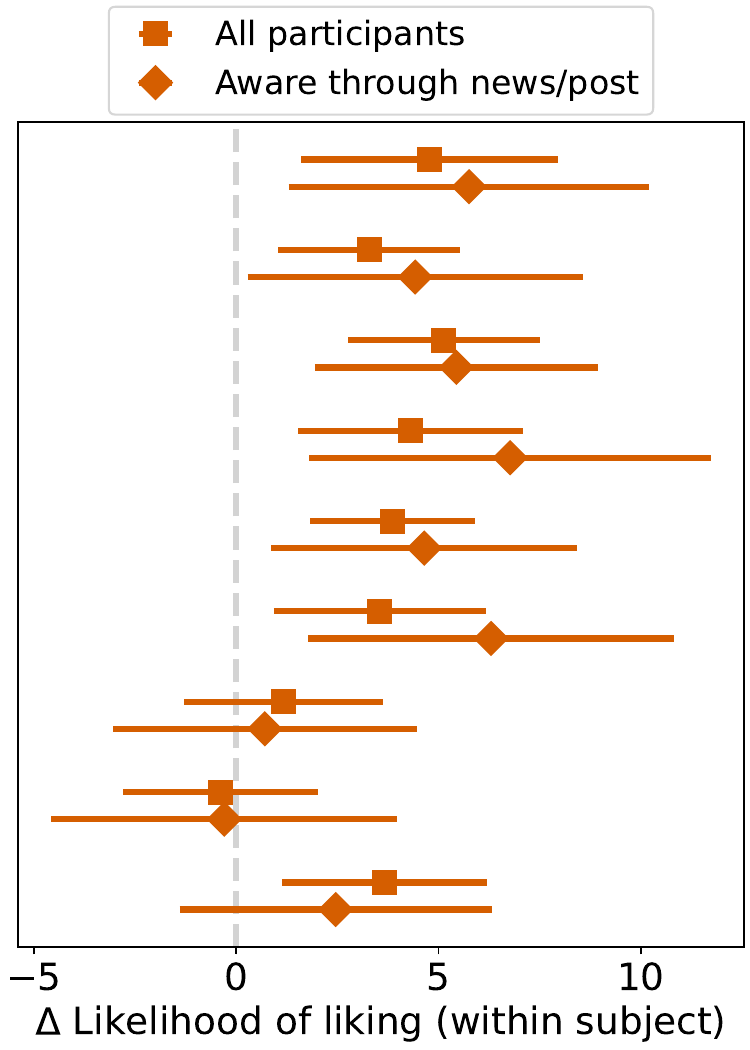}}
    \caption{Participants' self-rated likelihood of liking posts on high-reputational-risk and low-reputational-risk topics if their likes were public or private ($N=$ \num{203}). (a)~Mean likelihood of liking (0–100 scale) for each topic when likes were public or private (group means). (b)~Mean within-subject change ($\Delta$) in likelihood of liking between private and public conditions, plotted for all participants and for the subset who were aware of the platform's policy change through news or posts. Positive values indicate greater liking when likes were private. The error bars represent 95\% CIs.}
    \label{fig:like_tendency}
\end{figure}

\subsubsection*{Average likelihood of liking posts}
The average participants' self-rated likelihood of liking posts (\ie, group means) across high-reputational-risk and low-reputational-risk topics under public and private visibility conditions are presented in \cref{fig:like_tendency}a. For all high-reputational-risk topics, we find that the average self-rated willingness to like related posts is slightly higher under private visibility than under public visibility. For instance, willingness increases (i) from $35.481$ ($t(202.0)=14.339$, $p<0.001$; 95\% CI: $[30.602, 40.360]$) to $40.260$ ($t(202.0)=15.712$, $p<0.001$; 95\% CI: $[35.208, 45.313]$) for posts related to abortion; (ii) from $34.382$ ($t(202.0)=13.219$, $p<0.001$; 95\% CI: $[29.253, 39.510]$) to $37.663$ ($t(202.0)=14.419$, $p<0.001$; 95\% CI: $[32.513, 42.813]$) for posts related to BLM movement; (iii) from $38.567$ ($t(202.0)=15.861$, $p<0.001$; 95\% CI: $[33.772, 43.361]$) to $43.699$ ($t(202.0)=18.045$, $p<0.001$; 95\% CI: $[38.924, 48.474]$) for posts related to immigration \& border security; (iv) from  $46.971$ ($t(202.0)=18.863$, $p<0.001$; 95\% CI: $[42.061, 51.881]$) to $51.275$ ($t(202.0)=20.279$, $p<0.001$; 95\% CI: $[46.290, 56.261]$) for posts related to corruption in the government; (v) from $29.473$ ($t(202.0)=11.747$, $p<0.001$; 95\% CI: $[24.526, 34.420]$) to $33.333$ ($t(202.0)=13.159$, $p<0.001$; 95\% CI: $[28.338, 38.327]$) for posts related to transgender rights; and (vi) from $41.942$ ($t(202.0)=16.462$, $p<0.001$; 95\% CI: $[36.918, 46.966]$) to $45.491$ ($t(202.0)=18.226$, $p<0.001$; 95\% CI: $[40.569, 50.412]$) for posts related to gun control. However, the confidence intervals for the two conditions significantly overlap across these topics, indicating that making liker identities anonymous does not substantially alter participants' self-reported willingness to engage with controversial content.

For low-reputational-risk topics, we observe an even larger overlap between the public and private conditions than for high-reputational-risk topics, suggesting that making likes private has minimal impact on engagement with non-controversial content. For example, the mean likelihood of liking posts related to art is $57.572$ ($t(202.0)=25.271$, $p<0.001$; 95\% CI: $[53.080, 62.064]$) in the public mode and $58.738$ ($t(202.0)=25.607$, $p<0.001$; 95\% CI: $[54.215, 63.261]$) in the private mode. For charities working to end hunger, the means are $54.471$ ($t(202.0)=21.898$, $p<0.001$; 95\% CI: $[49.566, 59.375]$) in the public mode and $54.081$ ($t(202.0)=21.675$, $p<0.001$; 95\% CI: $[49.161, 59.000]$) in the private mode. Additionally, for pop culture, we observe a pattern similar to that we see in high-reputational-risk topics: participants' self-rated willingness slightly increases from $49.586$ ($t(202.0)=21.464$, $p<0.001$; 95\% CI: $[45.031, 54.141]$) to $53.248$ ($t(202.0)=22.976$, $p<0.001$; 95\% CI: $[48.679, 57.818]$), while this increase remains limited and does not indicate a substantial effect of private likes.

Notably, we find that the average participants' self-rated likelihood of liking posts across all topics center around 50 out of 100, suggesting a generally moderate inclination to engage with content through likes. Meanwhile, as shown in Fig.~\ref{fig:like_tendency}a, the likelihood of liking posts related to high-reputational-risk topics remains consistently lower than that for low-reputational-risk topics, regardless of the visibility condition.

\subsubsection*{Within-subject change in likelihood of liking posts}
At the group level, we find no evidence that private visibility significantly alters participants' self-reported willingness to engage with high-reputational-risk content. To capture more nuanced behavioral variation, we further examine the within-subject changes in participants' self-rated likelihood of liking posts between the public and private visibility conditions. This analysis allows us to assess whether individual participants, rather than aggregate group averages, exhibit shifts in engagement tendencies following the visibility change.

Specifically, we examine the within-subject changes by calculating the paired differences in each participant's likelihood of liking posts in the private condition relative to the public condition across high- and low-reputational-risk topics. As shown in Fig.~\ref{fig:like_tendency}b, we find that, for high-reputational-risk topics, the within-subject changes in participants' self-rated likelihood of liking posts consistently show small but statistically significant increases from public condition to private condition across abortion (mean $=4.780$, $t(202.0)=2.973$, $p=0.003$; 95\% CI: $[1.609, 7.950]$), BLM movement (mean $=3.281$, $t(202.0)=2.878$, $p=0.004$; 95\% CI: $[1.034, 5.529]$), immigration \& border security (mean $=5.133$, $t(202.0)=4.269$, $p<0.001$; 95\% CI: $[2.762, 7.503]$), corruption in the government (mean $=4.304$, $t(202.0)=3.062$, $p=0.002$; 95\% CI: $[1.533, 7.076]$), transgender rights (mean $=3.860$, $t(202.0)=3.750$, $p<0.001$; 95\% CI: $[1.830, 5.889]$), and gun control (mean $=3.549$, $t(202.0)=2.668$, $p=0.008$; 95\% CI: $[0.926, 6.172]$). Across low-reputational-risk topics, the within-subject changes for art (mean $=1.166$, $t(202.0)=0.937$, $p=0.350$; 95\% CI: $[-1.289, 3.622]$) and for charities working to end hunger (mean $=-0.390$, $t(202.0)=-0.320$, $p=0.749$; 95\% CI: $[-2.790, 2.010]$) are not statistically significant. However, for pop culture, the within-subject change is significantly positive (mean $=3.663$, $t(202.0)=2.861$, $p=0.005$; 95\% CI: $[1.138, 6.187]$). This suggests, for all participants, the within-subject changes are generally negligible across low-reputational-risk topics, with the exception of pop culture, which may also trigger mild self-censorship.

Additionally, we conduct a subset analysis for media-informed participants who learned about the policy change through news coverage or posts on \X. Given that the publication of news items or posts related to policy are time-sensitive, these media-informed participants were likely active online and aware of the visibility change shortly after its implementation. Consequently, focusing on media-informed participants can help us examine a relatively long-term effect compared to those who only learned about the policy change during the survey. \cref{fig:like_tendency}b shows that the within-subject changes among media-informed participants are significantly positive and consistent with those observed across all participants for high-reputational-risk topics: abortion (mean $=5.753$, $t(91.0)=2.564$, $p=0.012$; 95\% CI: $[1.296, 10.210]$), BLM movement (mean $=4.420$, $t(91.0)=2.122$, $p=0.037$; 95\% CI: $[0.283, 8.558]$), immigration \& border security (mean $=5.438$, $t(91.0)=3.091$, $p=0.003$; 95\% CI: $[1.943, 8.933]$), corruption in the government (mean $=6.766$, $t(91.0)=2.710$, $p=0.008$; 95\% CI: $[1.807, 11.726]$), transgender rights (mean $=4.645$, $t(91.0)=2.439$, $p=0.017$; 95\% CI: $[0.862, 8.427]$), and gun control (mean $=6.297$, $t(91.0)=2.768$, $p=0.007$; 95\% CI: $[1.778, 10.816]$). On the contrary, the within-subject changes among media-informed participants are statistically not significant across low-reputational-risk topics: art (mean $=0.704$, $t(91.0)=0.373$, $p=0.710$; 95\% CI: $[-3.046, 4.454]$), charities working to end hunger (mean $=-0.295$, $t(91.0)=-0.137$, $p=0.891$; 95\% CI: $[-4.572, 3.982]$), and pop culture (mean $=2.457$, $t(91.0)=1.265$, $p=0.209$; 95\% CI: $[-1.401, 6.316]$).

In summary, the within-subject changes consistently show that the participants' self-rated likelihood of liking posts significantly increases from the public to the private condition across all high-reputational-risk topics, while the within-subject changes are not statistically significant across non-controversial topics, particularly among media-informed participants. Moreover, as a robustness check, we repeat our analysis based on participants' self-rated likelihood of liking posts from specific high-reputational-risk and low-reputational-risk accounts (see \Cref{supp:suvery_accounts}). The account-based results are consistent with the topic-based findings. Taken together, these results suggest that making likes private indeed increases participants' willingness to engage with high-reputational-risk posts. However, these within-subject increases are not substantial enough to result in a meaningful change in the overall likelihood of liking posts, thereby not observable at the group level.

\subsubsection*{Heterogeneity among participants' demographics}

We further evaluate the within-subject changes in the self-rated likelihood of liking posts (\ie, $\Delta$likelihood) across participant's demographics, including age, sex, education, political leaning, living area, and household income (see details in \Cref{sec:exp_setup}). We find similar increases in $\Delta$likelihood to controversial topics between younger (mean $=3.607$, $t(93.0)=2.666$, $p=0.009$; 95\% CI: $[0.920, 6.294]$) and older participants (mean $=4.615$, $t(108.0)=3.909$, $p<0.001$; 95\% CI: $[2.275, 6.954]$), between male (mean $=3.685$, $t(122.0)=3.097$, $p=0.002$; 95\% CI: $[1.329, 6.040]$) and female participants (mean $=4.795$, $t(79.0)=3.586$, $p<0.001$; 95\% CI: $[2.133, 7.457]$), between left-leaning (mean $=4.683$, $t(105.0)=3.395$, $p<0.001$; 95\% CI: $[1.948, 7.418]$) and right-leaning participants (mean $=3.446$, $t(96.0)=3.279$, $p=0.001$; 95\% CI: $[1.360, 5.531]$), between urban (mean $=3.378$, $t(70.0)=2.194$, $p=0.032$; 95\% CI: $[0.307, 6.448]$) and non-urban participants (mean $=4.688$, $t(131.0)=4.312$, $p<0.001$; 95\% CI: $[2.537, 6.840]$). However, we find significant increases in $\Delta$likelihood to controversial topics for higher-educated (mean $=4.271$, $t(171.0)=4.284$, $p<0.001$; 95\% CI: $[2.303, 6.239]$) and higher-household-income participants (mean $=5.834$, $t(129.0)=4.785$, $p<0.001$; 95\% CI: $[3.422, 8.246]$), with no statistically significant increases for lower-educated (mean $=3.726$, $t(30.0)=1.836$, $p=0.076$; 95\% CI: $[-0.419, 7.870]$) and lower-household-income participants (mean $=2.009$, $t(72.0)=1.618$, $p=0.110$; 95\% CI: $[-0.466, 4.484]$). This pattern suggests that higher-educated and higher-income users engage in larger self-censorship, and thus exhibit larger increases in willingness of liking controversial posts when likes were private.

\section{Discussion}

Social media platforms provide important venues for users to interact with others and manage their own personas. Yet, these behaviors are shaped and constrained by the design and visibility of engagement features on the platforms~\cite{bucher2018affordances,marder2016like}. Recently, \X implemented a major design change on its engagement feature, making previously public likes private, which means that likers are only visible to post authors and themselves.
This change raises an important question: \emph{Did the private visibility encourage users to like sensitive or high-reputational-risk content more often compared to the public visibility?} 
Here, we conduct a comprehensive analysis from both the platform side and the user side. We find that the private visibility did not imply a substantial change in the number of likes for high-reputational-risk posts on \X. While we observe that users' self-reported willingness in liking controversial posts increased under the private condition, this increase did not result in a meaningful effect in behavior change.

Given that our analysis considers changes in the number of likes on high-reputational-risk posts within the four weeks following the implementation of private visibility, it is possible that some users were not yet aware of this change during this period. To address this concern, we conduct robustness check by shifting the post-change period to four months after the implementation of private-like policy. The results remain robust and consistent with our main findings (see \Cref{supp:did_robustness}). Additionally, we assume that relevant posts and news coverage were published shortly after the visibility update and examine the patterns of participants who learned about the change through these sources, \ie, media-informed participants. Our findings show that media-informed participants exhibited similar changes in willingness to like posts as other participants. Therefore, the lack of substantial changes in the number of likes to high-reputational-risk posts at the platform level is unlikely to be solely due to users' unawareness of the visibility update. Moreover, the consistent patterns between media-informed participants and participants who were aware of the visibility change through our survey suggest that the willingness changes are stable rather than an immediate reaction to awareness. 

We note that engagement is concentrated on social media platforms -- a small fraction of highly active users and/or automated agent accounts for a large share of engagement~\cite{ferrara2016rise,zhu2016attention,grinberg2019fake,baribi2024supersharers,shao2018spread}. This concentration can obscure individual-level behavioral trends, making subtle changes in engagement among ordinary users difficult to detect. In response, our robustness check still shows the null effect in the $\Delta$likes relative to accounts' like count baselines (see \Cref{supp:did_robustness}). 
However, engagement patterns are not only unequal but also strategically distributed. Users could maintain multiple or pseudonymous accounts that serve different social purposes, for instance, separating public-facing profiles from private or anonymous accounts~\cite{zhang2014anonymity,sharma2021identifying}. Users may use their secondary accounts to engage with content they perceive as controversial or reputationally risky, thereby decoupling their expressive behavior from their main identity. This behavioral adaptation can reduce the impact of visibility interventions.

Importantly, the divergence between users' self-reported willingness and observed behavior suggests an \emph{intention–behavior gap} in digital contexts. Previous research has shown that users are more willing to share controversial content under anonymous or private conditions~\cite{zhang2014anonymity}, and that liking intentions have positive impacts on liking behavior~\cite{chin2015facebook,levordashka2016s}. However, changing intentions does not guarantee behavior change, and there is a large  intention-behavior gap: a medium-to-large-sized change in intentions lead to only a small-to-medium-sized change in behavior~\cite{webb2006does,rhodes2012experimental}. Further evidence suggests that intentions get translated into action approximately one-half of the time~\cite{sheeran2016intention}. Consistent to this intention–behavior gap, the modest increase in willingness to like posts under private visibility on \X does not translate into substantial changes in actual engagement. Similar patterns have been observed in prior studies of platform design changes. For instance, in a case study of \X's character limit, previous research found that although post-intervention ``crammed'' tweets were longer, their syntactic and semantic characteristics remained largely unchanged, reflecting only superficial adjustments rather than meaningful behavioral shifts~\cite{gligoric2022anticipated}. Likewise, \X temporarily suspended the retweet function and prompted users to use the quote tweet function during 2020 \US presidential election. However, the suspension of the retweet did not encourage users to post more quotation tweets or longer comments~\cite{zhang2024effect}.

Our findings suggest that visibility interventions aimed solely at reducing reputational risk may be insufficient to meaningfully shift platform-wide behavior. Instead, such changes may need to be paired with complementary design levers -- such as algorithmic adjustments to content exposure, temporal interventions in feedback timing, and audience segmentation tools -- to influence engagement in more consequential ways. By \textit{algorithmic adjustments}, platforms can decouple private engagements from recommendation logic, reducing the fear of algorithmic surveillance. Similarly, to further reduce users' fear of surveillance, platforms could use \textit{temporal interventions} to alter when users receive engagement feedback, such as delaying or batching notifications about likes. By loosening the immediacy of feedback, these interventions weaken the sense that one's actions are being instantly observed and evaluated, thereby reducing the real-time reputational pressures associated with engaging with controversial content. Finally, using \textit{audience segmentation tools}, platforms could allow users to control which followers or sub-audiences can see their interactions, enabling more context-specific engagement without fully public exposure.

As with all research, ours is not free of limitations and offers opportunities for future research. First, we focus on aggregate engagement metrics (likes and reposts) and do not directly observe the individual-level interactions, which may limit understanding of nuanced behavioral responses. Future work with access to such data could examine how users perceive and respond to hidden liker identities, and how their behavior unfolds over time.
Second, engagement on \X is highly concentrated among a small fraction of highly active or automated accounts, which may obscure subtle effects among ordinary users and limit the detectable power of platform-level analyses. While this study conducts robustness check in Study 1 and subset analysis in Study 2 to mitigate this concern, future work could attempt to distinguish ordinary users from heavy users and bots to assess how making likes private affects different segments of the user base.
Third, our survey experiment relies on self-reported willingness to like posts, which may be subject to social desirability or hypothetical bias, and may not perfectly map onto actual behavior. Follow-up research could combine surveys with field interventions to more directly characterize the gap between what users say they would do under public vs. private like visibility and how they actually behave. Finally, our study focuses on a single platform, \X, and specific content types (\eg, high-reputational-risk and controversial posts). More research is needed to test the generalizability of our findings across other platforms, culture contexts, and engagement features.

\bibliographystyle{ACM-Reference-Format}
\bibliography{refs}

\appendix
\clearpage
\begin{center}
    \Large \textbf{Supplementary Materials}
    \vspace{1cm}
\end{center}

\renewcommand{\thetable}{S\arabic{table}}
\renewcommand{\thefigure}{S\arabic{figure}}
\renewcommand\thesection{S\arabic{section}}
\setcounter{figure}{0}
\setcounter{table}{0}
\setcounter{section}{0}

\section{Materials and Methods}
\label{supp:methods}

\subsection{Political influencers}
\label{supp:political_influencer}
\begin{figure}[ht]
    \centering
    \includegraphics[width=.7\linewidth]{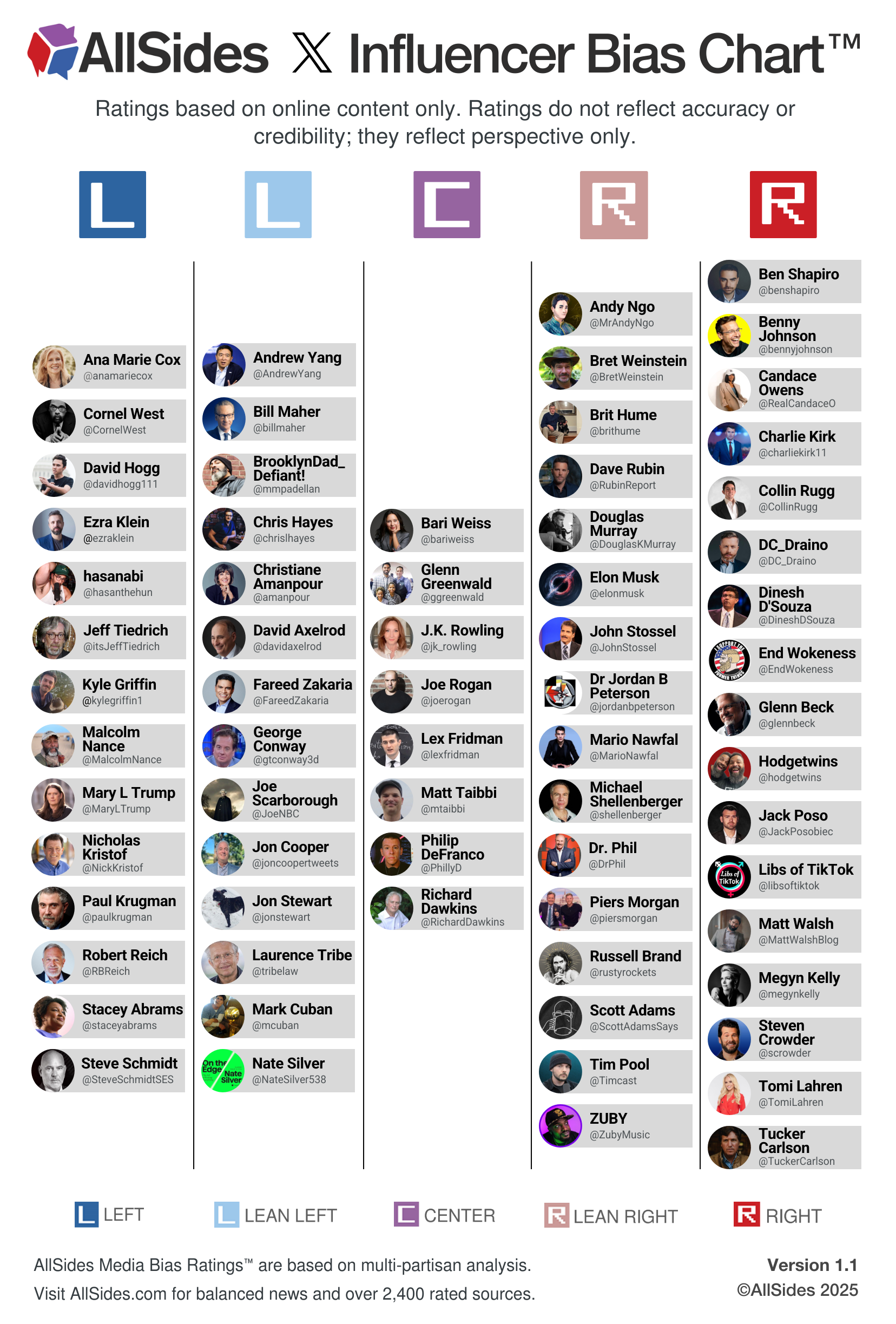}
    \caption{Influencer bias chart. The chart is sourced from AllSides at \url{https://www.allsides.com/media-bias/x-bias-chart}.}
    \label{fig:influencer_bias}
\end{figure}

We collect political influencers on \X based on the Influencer Bias Chart on AllSides (\cref{fig:influencer_bias}).

\clearpage
\subsection{User sample reweighting}
\label{supp:reweighting}

\begin{figure}[ht]
    \centering
    \captionsetup[subfloat]{font={bf, small}, skip=0pt, singlelinecheck=false, labelformat=simple, position=top}
    \subfloat[]{\includegraphics[width = .32\textwidth]{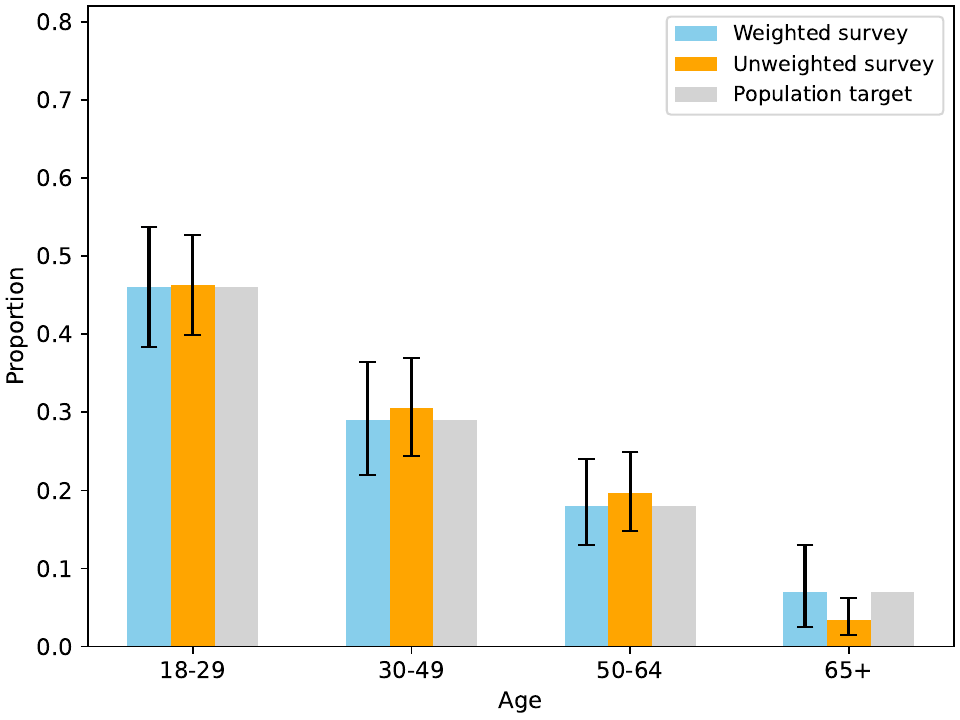}}
    \hfill
    \subfloat[]{\includegraphics[width = .32\textwidth]{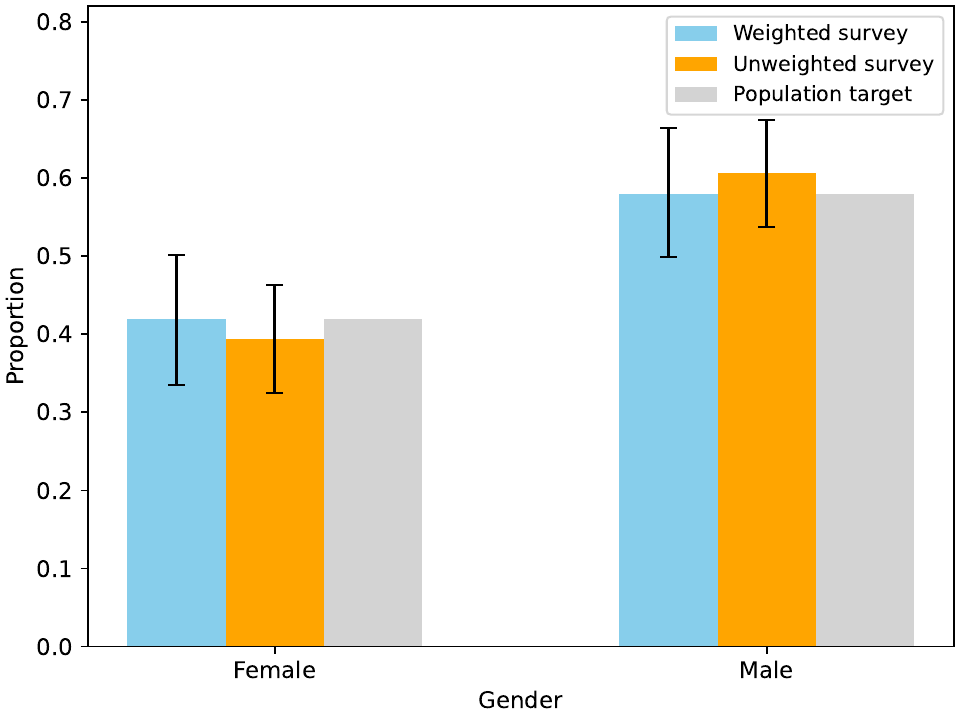}}
    \hfill
    \subfloat[]{\includegraphics[width = .32\textwidth]{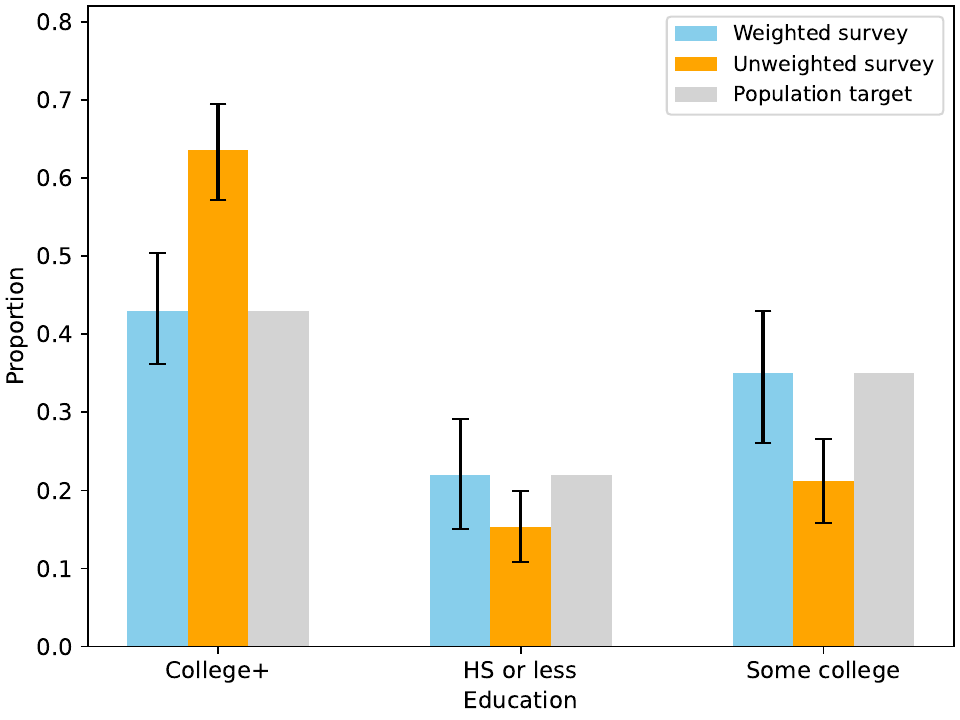}}

    \subfloat[]{\includegraphics[width = .32\textwidth]{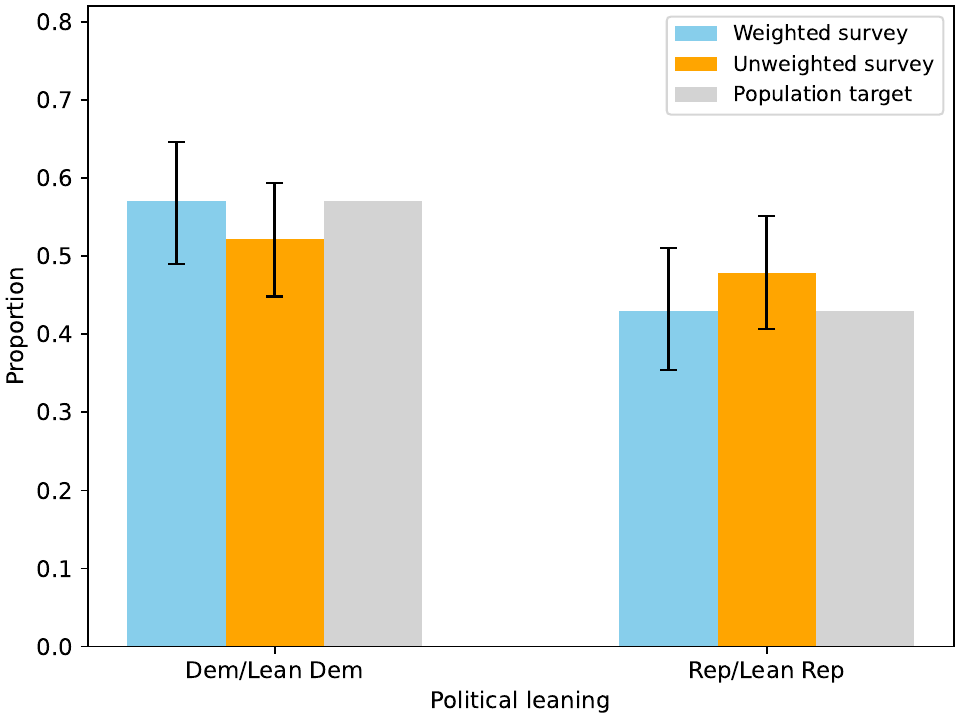}}
    \hfill
    \subfloat[]{\includegraphics[width = .32\textwidth]{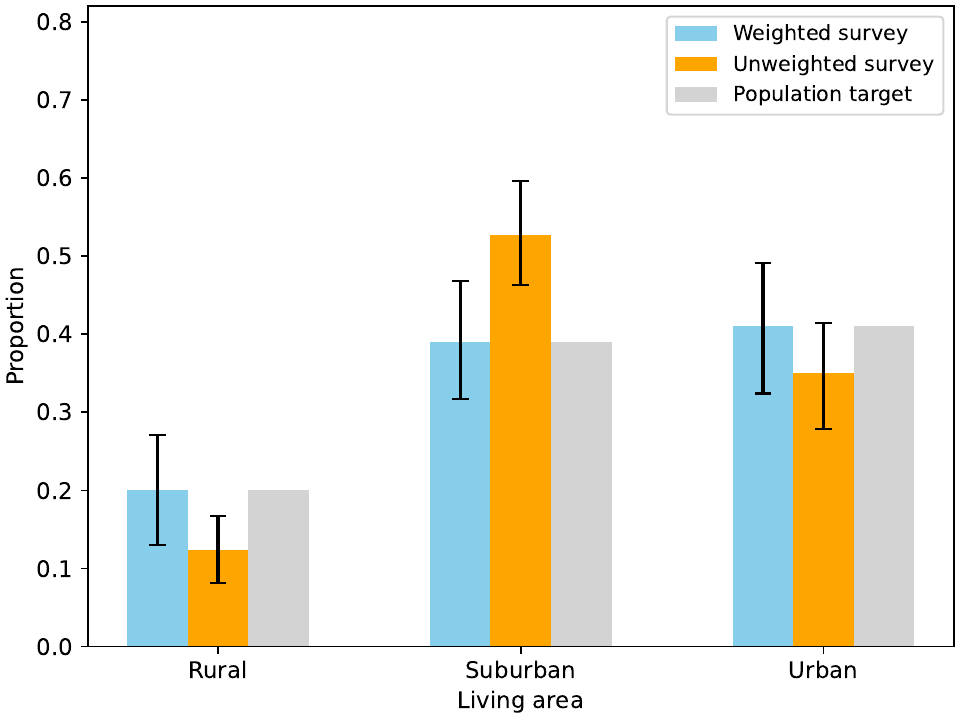}}
    \hfill
    \subfloat[]{\includegraphics[width = .32\textwidth]{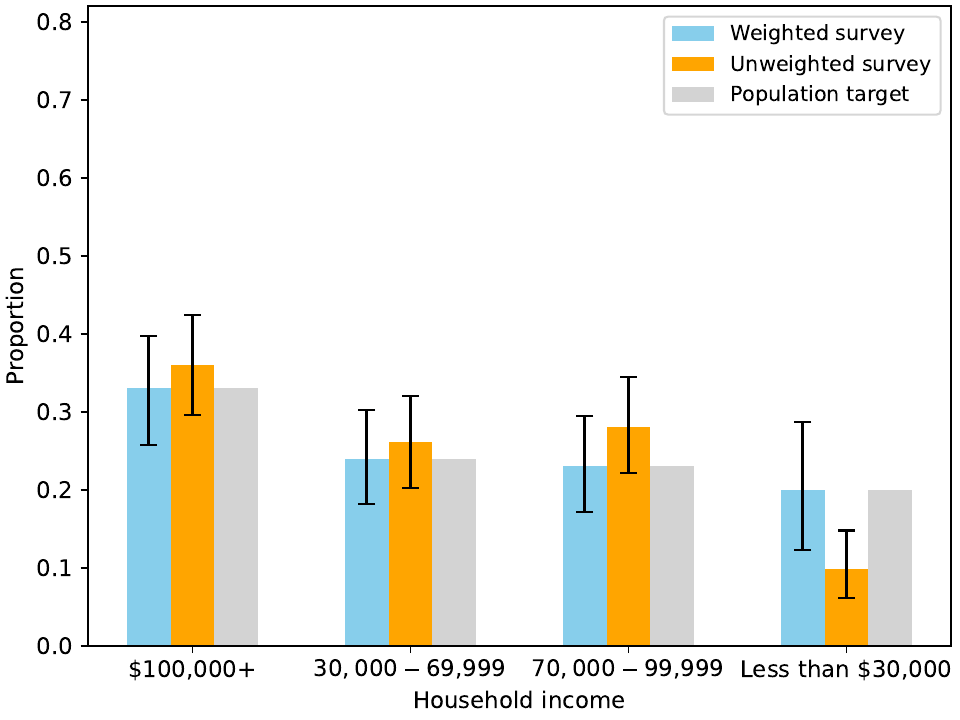}}
    
    \caption{Demographics of participants. (a)~The age distribution of participants. (b)~The gender distribution of participants. (c)~The education distribution of participants. (d)~The political leaning distribution of participants. (e)~The living area distribution of participant. (f)~The household income distribution of participants. The error bars represent 95\% CIs that are estimated based on bootstrap ($n=500$).}
    \label{fig:demographic}
\end{figure}

We apply iterative proportional fitting algorithm to reweight the sample to match Pew's demographic margins. The algorithm is implemented based on the ipfn package in Python.\footnote{\url{https://github.com/Dirguis/ipfn}} \cref{fig:demographic} presents the weighted and unweighted demographics of participants. The weighted demographics align well with the demographic distribution of overall population on \X.

\clearpage
\section{Robustness Checks for DiD Estimations}
\label{supp:did_robustness}

\subsection{HonestDiD estimation}
\begin{figure}[ht]
    \centering
    \captionsetup[subfloat]{font={bf, small}, skip=0pt, singlelinecheck=false, labelformat=simple, position=top}
    \subfloat[]{\includegraphics[width = .23\textwidth]{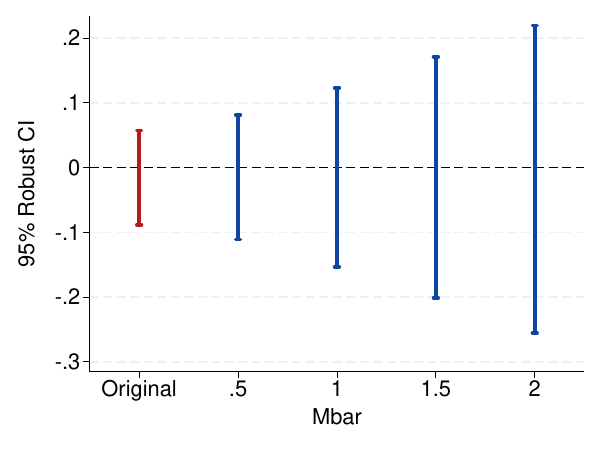}}
    \hfill
    \subfloat[]{\includegraphics[width = .23\textwidth]{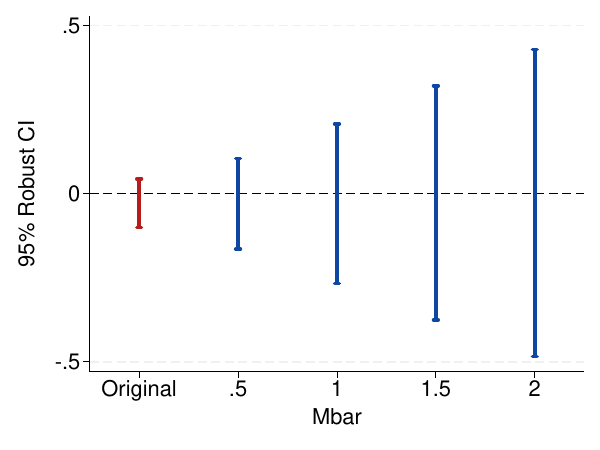}}
    \hfill
    \subfloat[]{\includegraphics[width = .23\textwidth]{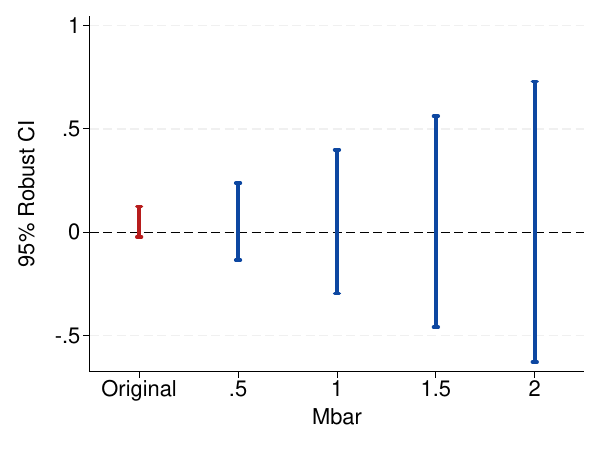}}
    \hfill
    \subfloat[]{\includegraphics[width = .23\textwidth]{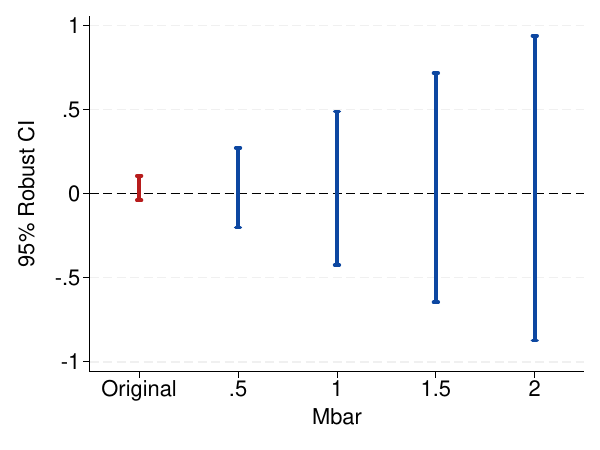}}

    \subfloat[]{\includegraphics[width = .23\textwidth]{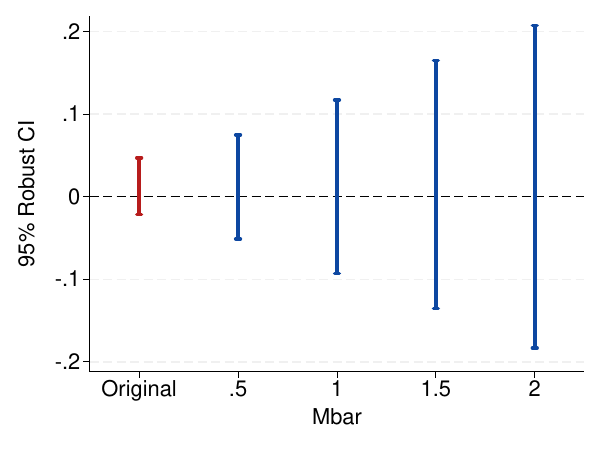}}
    \hfill
    \subfloat[]{\includegraphics[width = .23\textwidth]{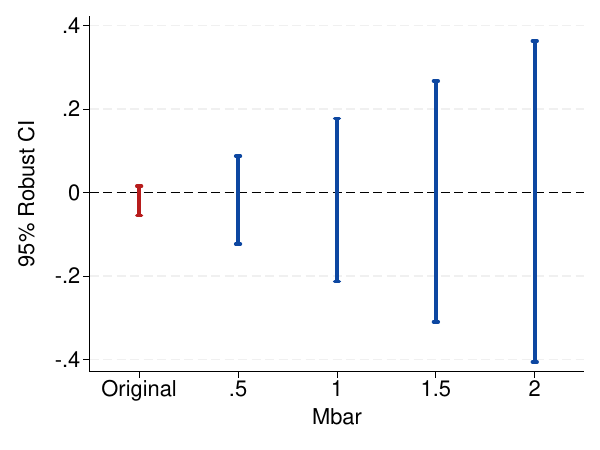}}
    \hfill
    \subfloat[]{\includegraphics[width = .23\textwidth]{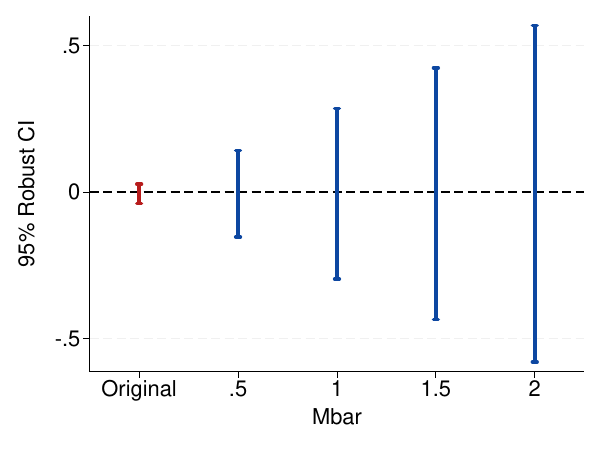}}
    \hfill
    \subfloat[]{\includegraphics[width = .23\textwidth]{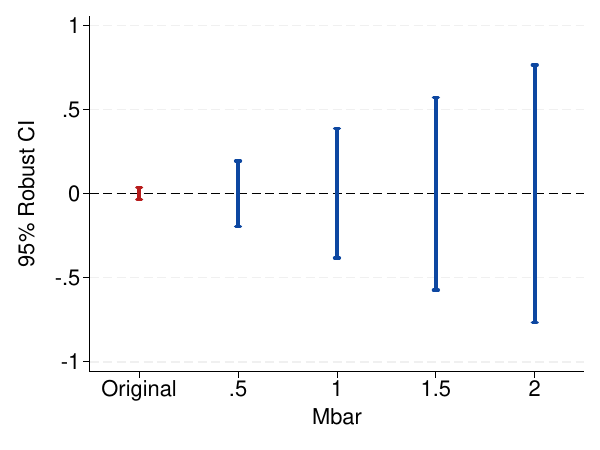}}
    
    \caption{The HonestDiD estimates for the post-change weekly periods. (a)--(d)~The HonestDiD estimates for the extra likes received by the high-reputational-risk accounts relative the low-reputational-risk accounts from week 1 to week 4 after the privacy change, respectively. (e)--(h)~The HonestDiD estimates for the extra engagement received in likes relative reposts from week 1 to week 4 after the privacy change, respectively.}
    \label{fig:honest_did}
\end{figure}

We use HonestDiD approach to calculate robust DiD estimates for the post-intervention weekly periods. All estimates remain robust and are not statistically significant (\cref{fig:honest_did}).

\clearpage
\subsection{Estimations of relative changes in likes}
Given that accounts can receive likes from stable sources -- such as their loyal followers, social bots, and high-activity users -- we use each account's average pre-intervention like count as a baseline, and examine the relative changes of likes over time across the treatment and control groups. Specifically, we remove two-sided 5\% outliers of like counts in the collected posts and calculate each account's baseline. Then, each post's like count is adjusted by subtracting the baseline of the corresponding account (\ie, $\Delta$likes). We adopt our DiD design and use linear regression to estimate $\Delta$likes over time. The estimation results are reported in \Cref{tab:did_diff}. All weekly DiD estimates and aggregated DiD estimate are not statistically significant (each $p>0.05$). Therefore, the changes in $\Delta$likes are also stable over time in the treatment group relative to the control group.

\begin{table*}[ht]
    \centering
    \setlength{\tabcolsep}{10pt}
    \caption{DiD estimates for $\Delta$likes.}
    \begin{tabular}{l*{4}{c}}
    \toprule
    &   DiD estimate   &   $t(152555)$  &   $p$   &   95\% CI\\
    \midrule
    Week $-8$    &  $81.438$    &   $t=0.205$   &   $p=0.838$   &   $[-698.652, 861.527]$\\
    Week $-7$    &  $697.438$    &   $t=1.745$   &   $p=0.081$   &   $[-85.864, 1480.741]$\\
    Week $-6$    &  $-608.996$    &   $t=-1.588$   &   $p=0.112$   &   $[-1360.729, 142.737]$\\
    Week $-5$    &  $-119.714$    &   $t=-0.306$   &   $p=0.759$   &   $[-885.871, 646.442]$\\
    Week $-4$    &  $178.747$   &   $t=0.456$   &   $p=0.649$   &   $[-589.94, 947.434]$\\
    Week $-3$    &  $318.774$  &   $t=0.8$ &   $p=0.424$   &   $[-462.005, 1099.554]$\\
    Week $-2$    &  $304.505$   &   $t=0.772$   &   $p=0.44$    &   $[-469.072, 1078.083]$\\
    Week $1$     &  $275.546$   &   $t=0.676$   &   $p=0.499$   &   $[-522.961, 1074.053]$\\
    Week $2$     &  $659.209$   &   $t=1.634$   &   $p=0.102$   &   $[-131.642, 1450.061]$\\
    Week $3$     &  $-301.77$   &   $t=-0.743$  &   $p=0.458$   &   $[-1098.123, 494.583]$\\
    Week $4$     &  $-287.568$  &   $t=-0.709$  &   $p=0.479$   &   $[-1083.011, 507.875]$\\
    Aggregated   &  $-6.913$    &   $t=-0.037$  &   $p=0.97$    &   $[-368.354, 354.528]$\\
    \bottomrule
    \end{tabular}
    \label{tab:did_diff}
\end{table*}

\clearpage
\subsection{Shift of post-treatment period}
Given that users might not be fully aware of the private likes within the first month (4 weeks) following the change, we shift our post-treatment period to four months later and collect posts from the identified accounts in the treatment and control groups over a four-week period since week 16 from likes made private. We repeat our analysis based on the new post-treatment period, and the estimation results are reported in \Cref{tab:did_long}. All weekly DiD estimates and aggregated DiD estimate are not statistically significant (each $p>0.05$). This means that the change to private likes did not increase engagement with reputationaly risky content in a long term.

\begin{table*}[ht]
    \centering
    \setlength{\tabcolsep}{10pt}
    \caption{DiD estimates for the number of likes with shifted post-treatment period.}
    \begin{tabular}{l*{4}{c}}
    \toprule
    &   DiD estimate   &   $z$  &   $p$   &   95\% CI\\
    \midrule
    Week $-8$    &$0.045$&   $z=1.225$   &   $p=0.221$   &   $[-0.026, 0.122]$\\
    Week $-7$    &$0.034$&   $z=0.909$   &   $p=0.363$   &   $[-0.038, 0.111]$\\
    Week $-6$    &$-0.01$&    $z=-0.3$   &   $p=0.764$   &   $[-0.076, 0.06]$\\
    Week $-5$    &$-0.028$&$z=-0.802$&$p=0.422$&$[-0.094, 0.042]$\\
    Week $-4$    &$-0.028$&$z=-0.793$&$p=0.428$&$[-0.095, 0.043]$\\
    Week $-3$    &$0.019$&$z=0.5$&$p=0.617$&$[-0.052, 0.094]$\\
    Week $-2$    &$0.003$&$z=0.086$&$p=0.931$&$[-0.065, 0.076]$\\
    Week $16$     &$0.02$&$z=0.522$&$p=0.602$&$[-0.052, 0.096]$\\
    Week $17$     &$0.036$&$z=0.942$&$p=0.346$&$[-0.037, 0.114]$\\
    Week $18$     &$-0.055$&$z=-1.549$&$p=0.121$&$[-0.121, 0.015]$\\
    Week $19$     &$-0.01$&$z=-0.266$&$p=0.79$&$[-0.078, 0.064]$\\
    Aggregated   &$-0.006$&$z=-0.37$&$p=0.711$&$[-0.038, 0.027]$\\
    \bottomrule
    \end{tabular}
    \label{tab:did_long}
\end{table*}

\clearpage
\section{Likelihood of Liking Across Accounts}
\label{supp:suvery_accounts}
As a robustness check, we evaluate participants' likelihood of liking posts based on account type alone, without exposure to specific posts. The survey includes (i) three categories of high-reputational-risk accounts: one SPLC extremist account, four politician accounts (two left-leaning and two right-leaning), and an account posting adult content; and (ii) two categories of low-reputational-risk accounts: one account on science \& technology and another account on charity. 

\begin{figure}[ht]
    \centering
    \captionsetup[subfloat]{font={bf, small}, skip=0pt, singlelinecheck=false, labelformat=simple, position=top}
    \subfloat[]{\includegraphics[width = .69\textwidth]{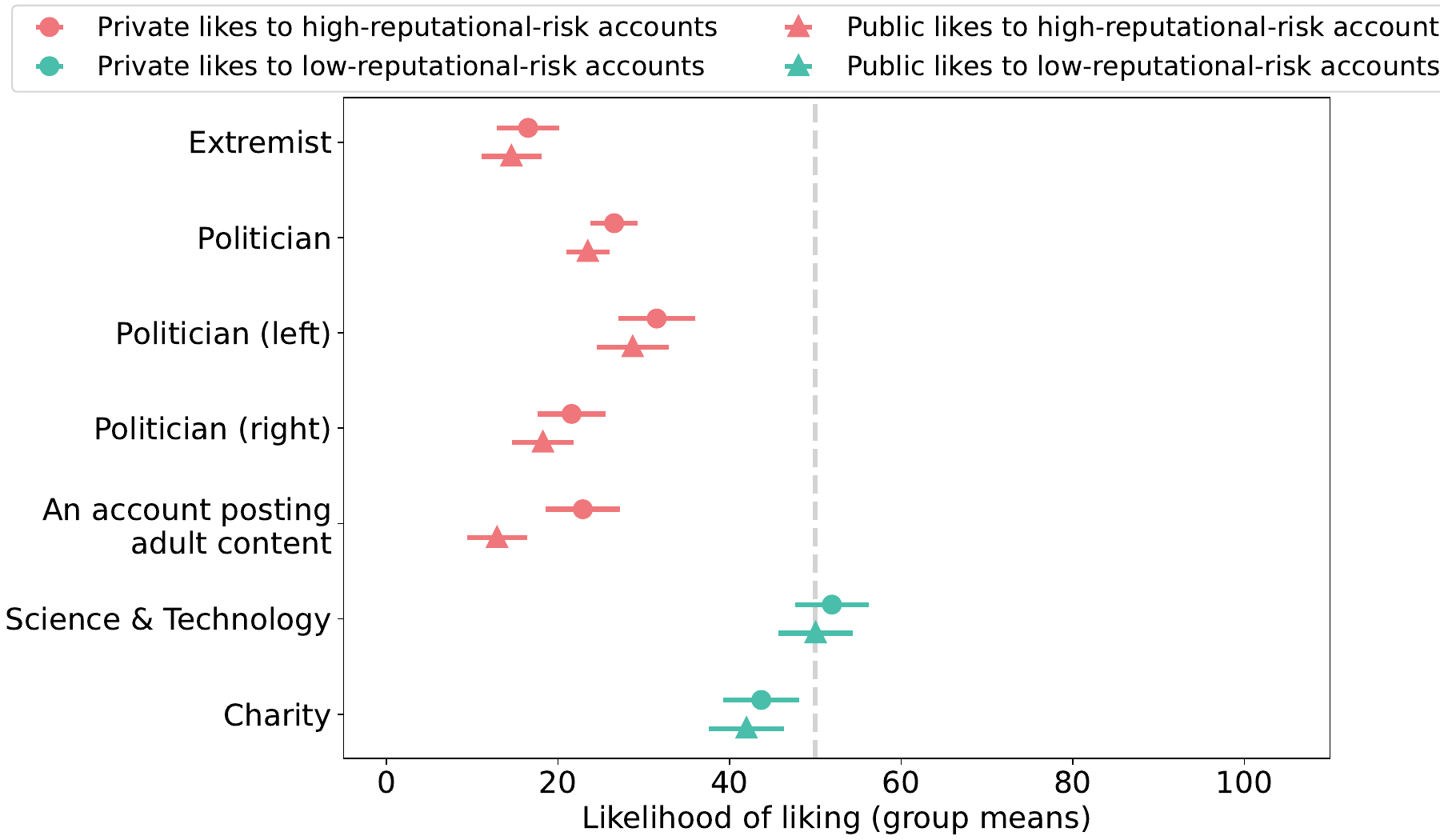}}
    \hfill
    \subfloat[]{\includegraphics[width = .288\textwidth]{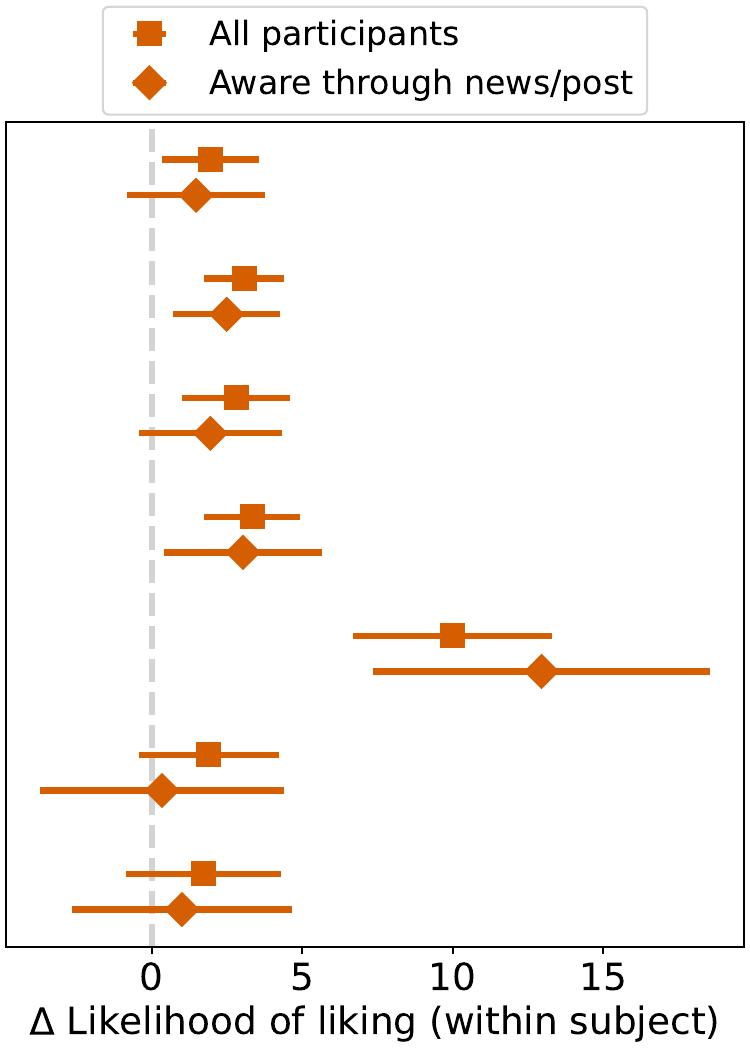}}

    \caption{Participants' self-rated likelihood of liking posts from high-reputational-risk and low-reputational-risk accounts if their likes were public or private ($N=$ \num{203}). (a)~Mean likelihood of liking (0–100 scale) for each account category when likes were public or private (group means). (b)~Mean within-subject change ($\Delta$) in likelihood of liking between private and public conditions, plotted for all participants and for the subset who were aware of the platform's policy change through news or posts. Positive values indicate greater liking when likes were private. The error bars represent 95\% CIs.}
    \label{fig:like_tendency_handles}
\end{figure}

We find that the mean likelihood of liking posts from the adult-content account increases significantly under private compared to public visibility (publich: mean $=12.898$, $t(202.0)=7.292$, $p<0.001$; 95\% CI: $[9.410, 16.386]$ vs. private: mean $=22.899$, $t(202.0)=10.373$, $p<0.001$; 95\% CI: $[18.546, 27.252]$), whereas the likelihood increases for the SPLC extremist (publich: mean $=14.569$, $t(202.0)=8.273$, $p<0.001$; 95\% CI: $[11.096, 18.042]$ vs. private: mean $=16.517$, $t(202.0)=8.958$, $p<0.001$; 95\% CI: $[12.881, 20.152]$) and politician accounts (public: mean $=23.478$, $t(202.0)=18.400$, $p<0.001$; 95\% CI: $[20.962, 25.994]$ vs. private: mean $=26.546$, $t(202.0)=18.882$, $p<0.001$; 95\% CI: $[23.774, 29.318]$) are modest and statistically non-significant (\cref{fig:like_tendency_handles}a). Additionally, the changes in the mean likelihood of liking posts are negligible for science \& technology (public: mean $=50.030$, $t(202.0)=22.943$, $p<0.001$; 95\% CI: $[45.730, 54.330]$ vs. private: mean $=51.927$, $t(202.0)=23.868$, $p<0.001$; 95\% CI: $[47.637, 56.217]$) and charity accounts (public: mean $=41.977$, $t(202.0)=18.797$, $p<0.001$; 95\% CI: $[37.573, 46.380]$ vs. private: mean $=43.701$, $t(202.0)=19.418$, $p<0.001$; 95\% CI: $[39.263, 48.138]$).

Furthermore, we compare the within-subject changes in the likelihood of liking posts across account types from pubic to private visibility (\cref{fig:like_tendency_handles}b). We find that, for high-reputational-risk accounts, the within-subject changes in participants’ self-rated likelihood of liking posts consistently show small but statistically significant increases from public condition to private condition across SPLC extremist (mean $=1.948$, $t(202.0)=2.396$, $p=0.017$; 95\% CI: $[0.345, 3.550]$), politicians (mean $=3.068$, $t(202.0)=4.527$, $p<0.001$; 95\% CI: $[1.732, 4.404]$), and adult-content account (mean $=10.002$, $t(202.0)=5.974$, $p<0.001$; 95\% CI: $[6.701, 13.303]$). However, the within-subject changes in participants’ self-rated likelihood of liking posts are statistically not significant across low-reputational-risk accounts: science \& technology account (mean $=1.897$, $t(202.0)=1.608$, $p=0.109$; 95\% CI: $[-0.430, 4.225]$) and charity account (mean $=1.724$, $t(202.0)=1.320$, $p=0.188$; 95\% CI: $[-0.851, 4.299]$).

Overall, these account-based findings are consistent with the topic-based results in our main analysis.

\end{document}